\documentclass[preprint,12pt]{elsarticle}

\usepackage{amssymb}

\newcommand\Rey{\mbox{\textit{Re}}}  
\newcommand\Ha{\mbox{\textit{Ha}}}  
\newcommand\Pran{\mbox{\textit{Pr}}} 
\newcommand\Nu{\mbox{\textit{Nu}}} 
\newcommand\N{\mbox{\textit{N}}} 
\newcommand\Gras{\mbox{\textit{Gr}}}  
\newcommand\Fr{\mbox{\textit{Fr}}}  

\newcommand{\dt}{\Delta t}

\newcommand{\bm}[1]{\mbox{\boldmath$#1$\unboldmath}}

\begin{document}

\begin{frontmatter}



\title{Thermal convection in a liquid metal battery}


\author{Yuxin Shen        and
        Oleg Zikanov\corref{cor1}}

\address{University of Michigan - Dearborn, 48128, MI, USA
              }
 \cortext[cor1]{Corresponding author, zikanov@umich.edu}

\begin{abstract}
Generation of thermal convection flow in the liquid metal battery, a device recently proposed as a promising solution for the problem of the short-term energy storage, is analyzed using a numerical model. It is found that convection caused by Joule heating of electrolyte during charging or discharging is virtually unavoidable. It  exists in laboratory prototypes larger than a few cm in size and should become much stronger in larger-scale batteries. The phenomenon needs further investigation in view of its positive (enhanced mixing of reactants) and negative (loss of efficiency and possible disruption of operation due to the flow-induced deformation of the electrolyte layer) effects.

\end{abstract}


\end{frontmatter}


\section{Introduction}
\label{sec:intro}
The work reported in this paper is motivated by the recent attempts to develop a commercially viable liquid metal battery, a device for large-scale, short-term, stationary energy storage \cite{Patent:2012,Wang:2014}. The battery, originally proposed in 1960s (see, e.g., \cite{Patent:1966}), is now a subject of renewed attention as a promising solution of the problem of intermittency of energy supply from wind and solar sources \cite{Kim:2013}. 

The energy stored in a liquid metal battery is the difference between the Gibbs free energy of a free light metal (e.g. Na, Li, or Mg) and of the same metal in a compound with a heavy metal (e.g. Bi, Sb, or PbSb). The processes of charging or discharging the battery correspond, respectively, to electrochemical reduction of the light metal from the compound or forming the compound. The reactions happen entirely in liquid state at the interfaces between the metals and a molten-salt electrolyte, which separates the metals from each other, immiscible with them, and is conductive to the ions of the light metal. An example of the electrolyte is LiF-LiCl-LiI, which can be used in a battery operating at temperatures about 450$^{\circ}$C with Li as a light metal  and SbPb  as a heavy metal \cite{Wang:2014}.

\begin{figure}
\centering
 \includegraphics[width=0.55\textwidth]{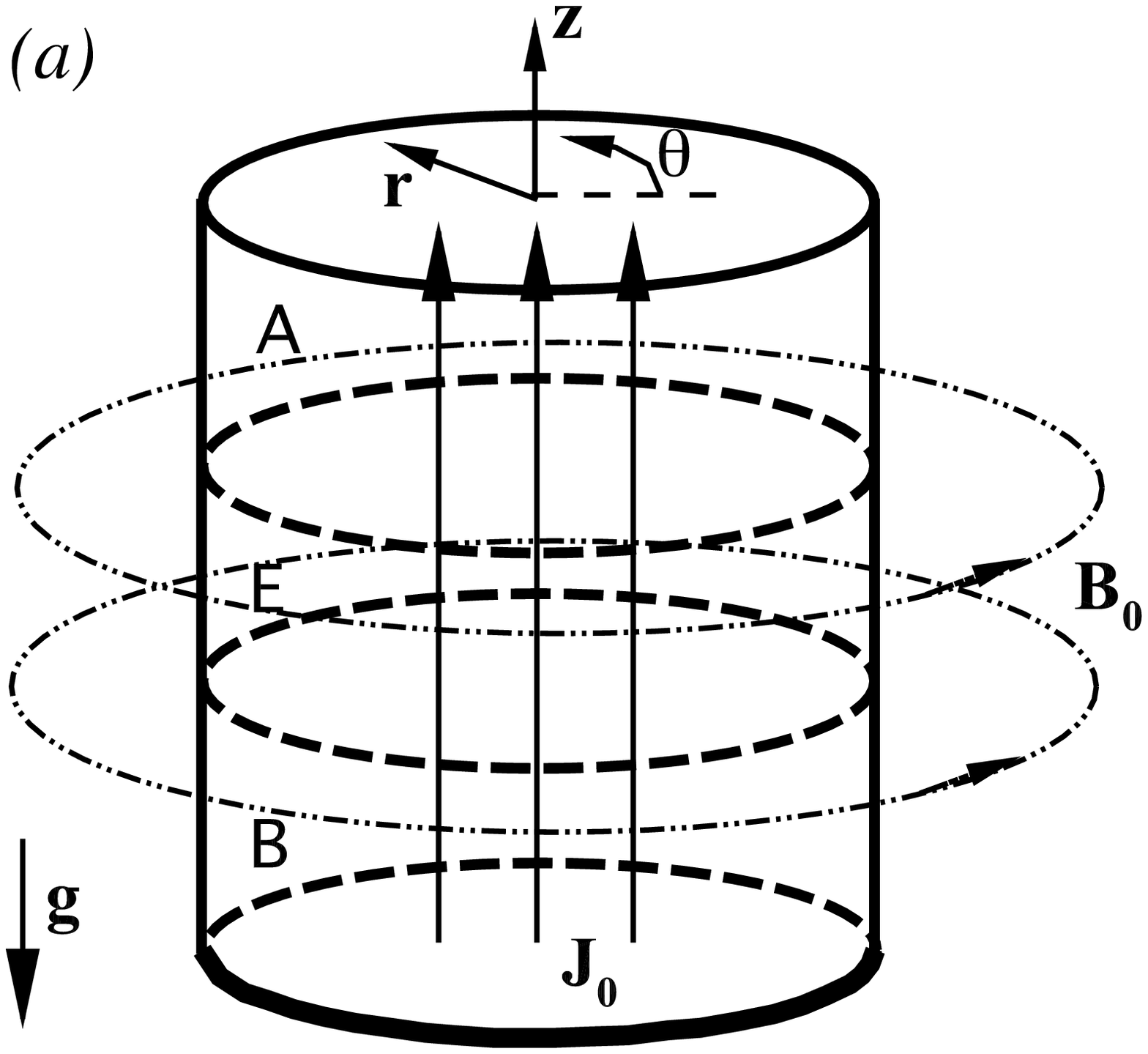}\includegraphics[width=0.45\textwidth]{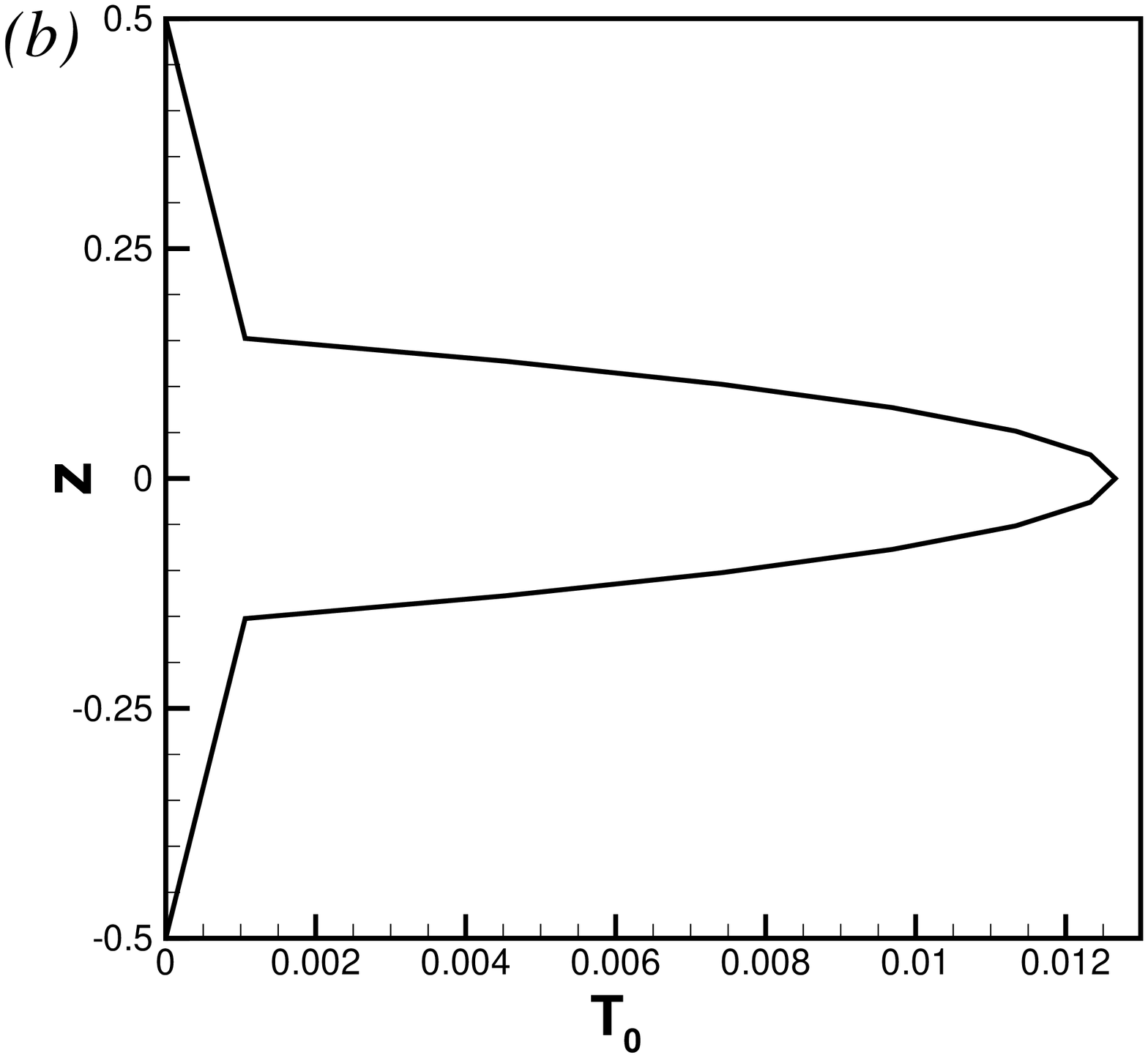}
 \caption{\emph{(a)}, Model geometry of the battery's model and the coordinate system. Three liquid layers \textsf{B},  \textsf{E}, and \textsf{A} fill a cylindrical cavity. During the charging and discharging processes the imposed uniform vertical electrical current of density $\bm{J}_0=J_0\bm{e}_z$ generates internal Joule heating and the azimuthal magnetic field $\bm{B}_0=B_0\bm{e}_{\theta}$. 
 \emph{(b)}, Typical distribution of non-dimensional temperature in the model battery in the absence of convection-generated flow (see section \ref{sec:model} for an explanation).}
\label{fig:geom}       
\end{figure}

The possible combinations of materials and features of the currently pursued battery designs are discussed, for example, in \cite{Kim:2013} and \cite{Wang:2014}. Important for us is that an operating battery can be viewed, in a simplified way, as a system schematically represented in Fig.~\ref{fig:geom}a, i.e., as a cylindrical box filled with three layers: the bottom layer \textsf{B} containing mixture of the heavy metal and the compound, the top layer \textsf{A} containing the light metal, and the electrolyte layer \textsf{E} in the middle. The system is stably stratified, with the densities of the materials satisfying $\rho_{B}>\rho_{E}>\rho_{A}$.  During the charging and discharging, strong (about 1 A/cm$^2$) electric current flows between the top and bottom walls serving as current collectors. The sidewalls are electrically insulating. 
 
The starting point of our work is the recognition that the system can experience hydrodynamic instabilities leading to flows in all three layers. The flows would significantly affect the operation of the battery, with implications, which can be both positive (faster electrochemical reactions due to enhanced mixing in the layer \textsf{B}) and negative (deformation of interfaces leading to nonuniform reaction rates and, in extreme cases, short circuit between the layers \textsf{A} and \textsf{B}). 

The instabilities can be considered as an essential part of the general scale-up problem, i.e. the problem of transition from small laboratory prototypes to large, commercially viable devices. It appears inevitable that increased size would lead to new instabilities, which would change the hydrodynamics of the system and affect its operation. 

The hydrodynamics of a liquid metal battery is largely unexplored, but general physical reasoning and preliminary studies suggest existence of several distinct instability mechanisms. One is the Tayler instability, the special case of the non-axisymmetric pinch-type instability in a column of an electrically conducting fluid with axial electric current. The instability, traditionally considered in the astrophysical context, has been recently analyzed for single-metal columns representing the battery in a drastically simplified way \cite{Seilmayer:2012,Stefani:2011,Weber:2014,Weber:2015}. It has been found that the instability threshold is predominantly determined by one non-dimensional parameter, the Hartmann number
\begin{equation} \label{hartman}
\Ha\equiv B_0(R)R\sqrt{\frac{\sigma}{\rho \nu}}, 
\end{equation}
where $B_0(R)$ is the magnitude of the azimuthal component of the magnetic field created by the axial electrical current at the outer radius R of the column, and $\sigma$, $\rho$, $\nu$ are the electrical conductivity, density, and kinematic viscosity of the metal. In the simplest case of an infinite column with the base state having zero flow velocity and uniform vertical current, the instability first occurs when $\Ha$ exceeds $\Ha_{cr}\approx 22$ and has the form of 
exponential growth of perturbations with the azimuthal wavenumber $m=1$ and the axial wavelength about $0.8\pi R$ \cite{Rudiger:2007}. Estimates based on the physical properties of typical liquid metals show that the instability occurs in moderately sized batteries. For example, a column of Mg at 750$^{\circ}$C with axial current of 1 A/cm$^2$ is unstable at $R$ above approximately 25 cm. We should note that, as proposed in \cite{Stefani:2011}, the Tayler instability can be shifted to higher $Ha$ or even completely avoided via modifications of battery's design that alter the base magnetic field $B_0$. Furthermore, as found in the recent analysis \cite{Herreman:2015}, even when the instability occurs in moderately sized batteries, the amplitude of the developing flow is small. Its kinetic energy is insufficient to overcome the gravitational potential of the stably stratified three-layer system and, so, to cause rupture of the electrolyte layer. It is suggested in  \cite{Herreman:2015} that the Tayler instability in a typical battery first becomes dangerous at $R\sim 1.5$ m or even higher.

Existence of another instability mechanism is suggested by the analogy between the liquid metal batteries and the aluminum reduction cells, the devices in which aluminum is produced from oxide by the electrochemical Hall-H\'{e}roult process. In a simplified description, a reduction cell is a large (about 4 by 12 m horizontally), shallow rectangular bath filled  by a layer of molten aluminum at the bottom and a layer of molten salt electrolyte, with aluminum oxide dissolved in it, at the top. It is known in the industry that the cells may experience instability in the form of growing sloshing motions of the interface between the two layers. The instability is magnetohydrodynamic in nature, related to the large difference between the electrical conductivities of electrolyte and aluminum, and caused by the interaction between the magnetic field generated by external electrical conductors and the perturbations of electrical currents within the cell arising due to the local deformations of the interface \cite{Sneyd:1994,Davidson:1998,Zikanov:2000,Sun:2004}.

In a liquid metal battery, the electrical conductivity of electrolyte is about four orders of magnitude lower than the conductivities of both metals. Any non-uniformity of the thickness of layer \textsf{E} would cause significant redistribution of electric currents within the battery and, thus, change of Lorentz forces acting in the melts. Since no results concerning this phenomenon have been published yet, we have to consider the possibility of the additional Lorentz forces leading to such a multi-layer magnetohydrodynamic instability as hypothetical. We should also note that the analogy between the battery and the aluminum reduction cell is incomplete, primarily because of significant differences in geometry.

The compositional convection can be caused by spatial variations of the concentrations of the heavy metal and compound in the bottom layer. While likely to be a factor affecting the battery's operation, this mechanism is impossible to analyze at the moment due to complexity and uncertainty of the relevant phase diagrams. We should also mention the short-wave instabilities of either Marangoni or electrohydrodynamic nature that may appear at the interfaces between the layers and may play a role in the dynamics of the battery. Finally, as suggested recently in \cite{Weber:2015}, flow in a battery can be generated by the electrovortex instability near the current collectors in the top and bottom walls.

Yet another, virtually unavoidable instability mechanism is that of thermal convection caused by non-uniformity of the temperature field. There are two sources of the non-uniformity: heating of the walls applied to maintain the necessary operational temperature and the internal  Joule heating of the poorly conducting electrolyte. The wall heating is required in small-scale laboratory prototypes but may become unnecessary in larger cells, where the Joule heating will be sufficient or even excessive (so the battery will have to be cooled) for temperature maintenance. 
To our knowledge, the only published results on convection in liquid metal batteries are those of the experiments \cite{Kelley:2014}, in which a simplified system consisting of a single liquid metal layer heated from below is considered. The focus of this work is on the expected positive effect of convection as a mechanism of mixing in the bottom layer. 

Our paper presents the first study of the convection caused by the internal Joule heating in the electrolyte layer. 
We apply numerical simulations to determine the typical size of the battery, at which convection flows become inevitable, and to understand the nature and properties of the flow.

\section{Model and approach to analysis}\label{sec:modmet}
\subsection{Physical model and governing equations}
\label{sec:model}
Even in its simplified form illustrated in figure \ref{fig:geom}, a liquid metal battery is a complex electro-magneto-hydrodynamic system. Its dynamics is a result of combined action of many mechanisms, some of which are possibly unknown. A reasonable first step of the analysis is to explore the known mechanisms individually, using idealized systems, in each of which the action of a particular mechanism is isolated. This was done, for example, for the Tayler instability in \cite{Seilmayer:2012,Stefani:2011,Weber:2014,Weber:2015,Herreman:2015} by neglecting the layered nature of the system and heating, i.e. by considering a column of an isothermal liquid metal. 

Here, we analyze the isolated effect of thermal convection. The effects of the Tayler instability, compositional convection, and the Lorentz forces associated with deformation of interfaces, short-wave interfacial instabilities, and electrovortex instabilities are removed by simplifying assumptions. As we show later, the approach is partially justified by the fact that the convection already occurs in batteries of very small size, significantly smaller than the sizes, at which the Tayler and, likely, other instabilities become active.

The simplifying assumptions made in our model are as follows:
\begin{enumerate}
\item The sidewalls of the battery are assumed thermally and electrically perfectly insulating.
\item The top and bottom walls, which, in a real battery, serve as current collectors, are assumed to be perfectly electrically conducting and, therefore, modeled as equipotential surfaces. We also assume that the top and bottom boundaries are maintained at a constant temperature. This corresponds to an external heating or cooling arranged in such a way that, together with the internal Joule heating, it creates a desired steady operational temperature within the battery.
\item The layered nature of the system is only partially retained in our model. Specifically, the electrical and thermal conductivities are assigned distinct and realistic values in each layer. For the other physical properties, such as density, viscosity, specific heat, and thermal expansion coefficient, we use the values typical for the electrolyte in all three layers. This departure from reality allows us to focus the analysis on the convection caused by non-uniform internal Joule heating and affected by the magnetic field. It also allows us to use the Boussinesq approximation, which would be impossible at finite density differences between the layers \cite{convbook:2012}, and to apply the effective computational method described in section \ref{sec:method}.
\item The base state is that of zero flow velocity in all three layers and the electrical current of constant and uniform density 
\begin{equation}
\bm{J}_0=J_0\bm{e}_z, \:\:  J_0=const
\end{equation} 
flowing between the top and bottom walls (see figure \ref{fig:geom}a). The current generates the constant, uniform, and purely azimuthal magnetic field 
\begin{equation}
\bm{B}_0=B_0(r)\bm{e}_{\theta}=\frac{\mu_0J_0 r}{2}\bm{e}_{\theta},
\end{equation} 
where $\mu_0$ is the magnetic permeability of free space.
\item The interfaces between the layers are modeled as steady-state, horizontal, and impermeable surfaces. The effects related to the deformation of the interfaces are, thus, neglected in our model. This is justified by the fact that the actual battery systems are strongly density-stratified. The density of the light metal is, at least, two times smaller than the density of the electrolyte, which, in turn, is several times smaller than the density of the compound between the light and heavy metals. A further discussion of this approximation based on the results of our computations and supporting its validity is given in section \ref{sec:disc}.
\item The coupling between the flows in adjacent layers  via viscous shear stresses, heat transfer, pressure forces, and electromagnetic effects is included into the model.
The effect of the surface tension at the interfaces is assumed to be weak in comparison to these mechanisms and neglected.   
\item The fluid is assumed to be Newtonian, incompressible, electrically conducting and, apart from the stepwise changes of thermal and electrical conductivities at the interfaces, having constant physical properties.
\item We assume that the typical timescale of the evolution of convection flows is much smaller than the typical time of charging or discharging the battery. For this reason, the thicknesses and physical properties of the three layers are assumed constant.
\item The Boussinesq approximation is used to describe the convection effect.
\item The quasi-static approximation \cite{Davidson:2001} is used to evaluate the  electric current perturbations induced by the flow velocity and the Lorentz forces resulting from the interaction of these currents with the base magnetic field $\bf{B}$. The approximation is considered valid because the magnetic Reynolds number based on the typical velocity and length scales of the convection flow is expected to be small. The perturbations of the magnetic field and, thus, the mechanism leading to the Tayler instability are not considered.
\end{enumerate}

To non-dimensionalize the governing equations and boundary conditions, we use the physical properties of electrolyte: density $\rho_{E}$, kinematic viscosity $\nu_E$, electrical conductivity $\sigma_E$, thermal conductivity $\kappa_E$, specific heat $C_E$, and thermal expansion coefficient $\alpha_E$. The radius of the battery $R$ is used as the typical length scale. In order to derive the temperature and velocity scales, we need, first, to consider the internal Joule heating by the imposed current $\bm{J}_0$. In the electrolyte, the volumetric heating rate is
\begin{equation}
Q_0=\frac{J_0^2}{\sigma_E}.
\end{equation}
This constant will be used as the typical scale for the internal heating rate, which is equal to $\sigma_A^{-1}J^2$ and $\sigma_B^{-1}J^2$ in, respectively, layers \textsf{A}  and \textsf{B}. The typical scales of temperature and velocity can then be set to
\begin{equation}\label{scales}
\Delta T=\frac{Q_0R^2}{\kappa_E}=\frac{\sigma_E^{-1}J_0^2 R^2}{\kappa_E}, \:\: U=\sqrt{\alpha_E g \Delta T R},
\end{equation}
where $g$ is the gravity acceleration constant.
The typical time and pressure scales are $RU^{-1}$ and $\rho_EU^2$. For the density of the additional electrical currents induced by the flow in the imposed magnetic field $\bm{B}_0$, we use the Ohm's law to derive the typical scale $\sigma_E U B_0$. The corresponding scale for the electric potential is $UB_0R$. The imposed magnetic field $\bm{B}_0$ is scaled by its magnitude at the sidewall $B_0(R)=\mu_0J_0 R/2$.

The imposed temperature of the top and bottom walls is taken as the reference temperature and set to zero. The gravity force and the base-state Lorentz force $-J_0B_0\bm{e}_r$ resulting from the interaction between the imposed current $\bm{J}_0$ and the magnetic field $\bm{B}_0$ are removed from consideration by subtracting from the total pressure field the steady-state pressure distributions balancing these forces. 

The non-dimensional governing equations are:
\begin{eqnarray}
\hskip-15mm& & \frac{\partial \bm{u}}{\partial t} + (\bm{u}\cdot \nabla)\bm{u} = -\nabla p +  \frac{1}{\Rey}\nabla^2 \bm{u} + \frac{\Ha^2}{\Rey}(\bm{j}\times \bm{B}_0)+T\bm{e}_z,\label{nsviscous}\\
\hskip-15mm& & \nabla \cdot\bm{u}  =  0,\label{incompr}\\
\hskip-15mm& & \frac{\partial T}{\partial t} + \bm{u}\cdot \nabla T = \frac{1}{\Rey\Pran} [\nabla\cdot(\kappa \nabla T)+Q],\label{scalar} \\
\hskip-15mm& & \bm{j}=\sigma(-\nabla \phi +\bm{u}\times\bm{B}_0), \label{ohm}\\
\hskip-15mm& & \nabla\cdot (\sigma \nabla \phi)=\nabla\cdot(\sigma \bm{u}\times\bm{B}), \label{poiss}
\end{eqnarray} 
where $\bm{u}$, $T$, $\bm{j}$, $\phi$, and $p$ are the non-dimensional velocity, temperature, density of  electric currents induced by the flow,  the  electric potential associated with these currents, and the modified pressure. The equations are written for the entire flow domain shown in figure \ref{fig:geom}. The non-dimensional rate of Joule heat generation $Q$, thermal conductivity $\kappa$, and electrical conductivity $\sigma$ change discontinuously at the interfaces between the layers and are defined as:
\begin{eqnarray}\label{discont1}
Q^{-1}=\sigma & = & \left\{
  \begin{array}{lr}
 \sigma_B/\sigma_E   & \textrm{at } z<z_E-H_E/2, \\
    1  & \textrm{at } z_E-H_E/2\le z\le z_E+H_E/2,\\
    \sigma_A/\sigma_E  & \textrm{at } z>z_E+H_E/2 ,
  \end{array}
\right.\\
\label{discont2}
\kappa & = &\left\{
  \begin{array}{lr}
   \kappa_B/\kappa_E   &  \textrm{at } z<z_E-H_E/2 ,\\
    1  &  \textrm{at } z_E-H_E/2\le z\le  z_E+H_E/2,\\
     \kappa_A/\kappa_E   & \textrm{at } z>z_E+H_E/2 ,
  \end{array}
\right.
\end{eqnarray} 
where $z_E$ and $H_E$ stand for the location of the midplane and thickness of the electrolyte layer, so $z<z_E-H_E/2$, $z_E-H_E/2\le z\le  z_E+H_E/2$, and $z>z_E+H_E/2$ correspond to the layers \textsf{B}, \textsf{E}, and \textsf{A}, respectively.  

The boundary conditions are 
\begin{eqnarray}
\label{bc1} \hskip-15mm& &  T=0, \: \bm{u}=0, \: \phi=0 \:\: \textrm{at} \: z=0, H,\\
\label{bc2} \hskip-15mm& &  \frac{\partial T}{\partial r}=0, \: \bm{u}=0, \: \frac{\partial \phi}{\partial r}=0 \:\: \textrm{at} \: r=1.
\end{eqnarray}
Additional boundary conditions follow from our assumption of non-deformable, impermeable, horizontal interfaces, which implies zero vertical velocity at them:
\begin{equation}
\label{intbc}
u_z=0 \: \textrm{at} \: z=z_E\pm H_E/2.
\end{equation}

The non-dimensional parameters of the problem are the Reynolds number
\begin{equation}
\label{reynolds}
\Rey\equiv \frac{UR}{\nu_E}=\Gras^{1/2},
\end{equation}
where
\begin{equation}
\label{grashof}
\Gras\equiv \frac{\alpha_E g Q_0 R^5}{\kappa_E \nu_E^2}=\frac{\alpha_E g J_0^2 R^5}{\sigma_E \kappa_E \nu_E^2}
\end{equation}
is the Grashof number,
the Prandtl number
\begin{equation}
\label{peclet}
\Pran\equiv\frac{\nu_E \rho_EC_E}{\kappa_E},
\end{equation}
the  Hartmann number
\begin{equation}
\label{hartmann}
\Ha\equiv B_0(R)R\left(\frac{\sigma_E}{\rho_E\nu_E}\right)^{1/2}=\frac{\mu_0J_0}{2}R^2\left(\frac{\sigma_E}{\rho_E\nu_E}\right)^{1/2},
\end{equation}
and the geometric parameters, namely the aspect ratio $H$ of the battery, and the thickness $H_E$ and location $z_E$ of the electrolyte layer.

As the base state of the system, we consider the always existing mathematical solution with zero velocity and electric potential, constant modified pressure, and the piecewise-parabolic velocity profile $T_0(z)$ found as an analytical solution of 
\begin{eqnarray}
\frac{d}{dz}\left[\kappa(z)\frac{dT_0}{dz}\right] & = & -Q(z), \\
T_0(0)=T_0(H) & = & 0, 
\end{eqnarray}
where $\kappa(z)$ and $Q(z)$ are defined by (\ref{discont1}) and (\ref{discont2}). As an illustration, the profile of $T_0(z)$ at $z_E=0$, $H_E=0.304$ is shown in figure \ref{fig:geom}b.


\subsection{Approach to analysis}\label{sec:procedure}
The convection instability and the resulting flows are analyzed via direct numerical solution of the system of equations and boundary conditions (\ref{nsviscous}--\ref{intbc}). 
Each simulation starts with random low-amplitude ($\sim 10^{-5}$) perturbations of velocity and temperature added to the base state. The base state is deemed stable or unstable if, after initial adjustment, the energy of the perturbations, respectively, decays or grows over a long (at least a hundred
units) time period. In all the cases of growing perturbations, we have been able to identify sufficiently long periods of exponential growth, during which the growth rate
\begin{equation}
\label{gamma}
\gamma=\frac{1}{2E}\frac{dE}{dt}\approx \frac{1}{2E^T}\frac{dE^T}{dt}, 
\end{equation}
of the volume-averaged perturbation energy
\begin{equation}\label{perten}
E=\frac{1}{V}\int_V |\bm{u}|^2dv \:\: \textrm{or} \:\:  E^T=\frac{1}{V}\int_V (T-T_0)^2dv
\end{equation}
is constant within the second significant digit.

The growth of perturbations is followed into the stage of a fully developed flow characterized by nonlinear saturation. After that, the flow's evolution is computed for not less than 1000 time units. Flow statistics are accumulated and time-averaged during this period.

\subsection{Parameter range under investigation}\label{sec:parameters}
As we have already discussed, the convection instability is expected to appear in the process of scale-up of a battery from small laboratory prototypes to large commercial devices. The important questions are: at what size the convection first appears in a battery of a certain design, and how the strength of convection and its mixing and heat-transfer effects vary with the size. We choose to address the questions and, so, vary the non-dimensional parameters of the problem in the manner that corresponds to variation of the size of a battery of a given design. We also explore the effect of the thickness of the electrolyte layer. The physical properties of the melts, imposed electric current density,  the aspect ratio of the cell, and the location of the mid-plane of the electrolyte layer are assumed constant. This means that we are left with two independent parameters: the Grashof number $\Gras$ and the non-dimensional electrolyte thickness $H_E$. The other parameters defined in section \ref{sec:model} are either constant ($\Pran$, $H$, $z_E$, $\sigma_B/\sigma_E$, $\sigma_A/\sigma_E$, $\kappa_B/\kappa_E$, $\kappa_A/\kappa_E$) or functions of $\Gras$: the already mentioned $\Rey=\Gras^{1/2}$ and
\begin{equation}\label{hagra}
\Ha=\beta \Gras^{2/5},
\end{equation}
where the coefficient
\begin{equation}\label{beta}
\beta=\frac{\mu_0}{2}\left( \frac{J_0^2\sigma_E^9\kappa_E^4\nu_E^3}{\rho_E^5\alpha_E^4g^4}\right)^{1/10} 
\end{equation} 
(see (\ref{grashof}) and (\ref{hartmann})).

Cells of the aspect ratio $H=1$ and with the imposed electric current $J_0=1$ A/cm$^2$ \cite{Patent:2012,Kim:2013,Wang:2014} are considered. The choice of physical properties is more difficult. The available data on the properties of the high-temperature melts used in the batteries are incomplete and, generally, of low quality. Further uncertainty is created by the presence of the ions of the light metal in the electrolyte, mixture of the heavy metal and the compound in the bottom layer, and the fact that the related phase diagrams are poorly known. For these reasons, we select a set of physical properties, which is typical in the sense that it represents the properties of the battery materials on the level of the orders of magnitude, but does not correspond to any specific battery design. 

The well-documented properties of LiCl-KCL at about 450$^{\circ}$C are chosen for the electrolyte: $\rho_E=1.63\times 10^3$ kg/m$^3$, $\alpha_E=2.93\times 10^{-4}$ K$^{-1}$, $C_E=1.21 \times 10^3$ J/kg$\cdot$K, $\nu_E=0.71\times 10^{-6}$ m$^2$/s, $\kappa_E=0.42$ W/m$\cdot$K, $\sigma_E=170$ S/m (see \cite{Kim:2013,Williams:2006}). This gives 
\begin{equation}\label{values}
\Pran=3.33, \:\: \beta=1.05 \times 10^{-6}.
\end{equation}
The battery's radius is related to the Grashof number as
\begin{equation}
\label{estimateR}
R=\left(\Gras\frac{\sigma_E \kappa_E \nu_E^2}{\alpha_E g J_0^2 } \right)^{1/5}=6.60\times 10^{-4}\Gras^{1/5} \textrm{ [m]}.
\end{equation}

The ratios of electrical and thermal conductivities are set at 
\begin{equation}\label{ratios}
\frac{\sigma_B}{\sigma_E}=\frac{\sigma_A}{\sigma_E}=10^4, \: \frac{\kappa_B}{\kappa_E}=\frac{\kappa_A}{\kappa_E}=50,
\end{equation}
which  corresponds to typical high electrical and thermal conductivities of liquid metals.

\subsection{Numerical method}
\label{sec:method}
The evolution of a three-dimensional unsteady flow is calculated in the manner of direct numerical simulation using the method introduced in \cite{Krasnov:FD:2011} and extended to flows with cylindrical geometry in \cite{Zhao:2012,ZikanovJFM:2013} and flows with the combined effects of convection and magnetic field in \cite{ZikanovJFM:2013,Zhang:2014,Lv:2014,Liu:2015,Krasnov:2012}. The details of the method can be found in these references, while only the main principles and the new features introduced in this study are described here. 

The method is based on the second-order finite-difference discretization on a collocated, structured, non-uniform grid.  The spatial derivatives are evaluated with the use of velocity and current fluxes obtained by interpolation to staggered grid points (see \cite{Krasnov:FD:2011}). This makes the scheme nearly fully conservative in the sense that, in the non-dissipative limit, it perfectly conserves mass, momentum, electric charge, and internal energy, while the kinetic energy is conserved with the dissipative error of the third order. 

The time-discretization uses the standard projection method to find pressure and satisfy incompressibility, and the second order backward-difference-Adams-Bashfort scheme modified so that the temperature diffusion term is treated implicitly. Each time step is accomplished as a sequence of the following substeps:

\emph{(i)} Use the fields obtained at the previous time level $t^n$ to solve the potential equation and find the electric currents
\begin{eqnarray}
\nabla\cdot\left(\sigma \nabla \phi^{n}\right) & = & \nabla \cdot\left(\sigma \bm{u}^n \times \bm{B}_0\right), \label{step1}\\
\label{step2}
\bm{j}^n & = & \sigma \left(-\nabla \phi^n + \bm{u}^n \times \bm{e}_y\right),\\
\end{eqnarray}
and compute
\begin{eqnarray}
\label{step4} \bm{F}^n & = & -\bm{M}(\bm{u}^n,\bm{u}^n)+\frac{1}{\Rey} \nabla^2 \bm{u}^n + \frac{\Ha^2}{\Rey} \bm{j}^n\times \bm{B}_0 +T\bm{e}_z,\\
\label{step5}  G^n & = & -\nabla \cdot\left(T^n \bm{u}^n \right)  +\frac{Q}{\Rey \Pran},
\end{eqnarray}
where $\bm{M}(\bm{u}^n,\bm{u}^n)$ is the nonlinear term in divergence form.

\emph{(ii)} Find the intermediate velocity field $\bm{u}^*$ from:
\begin{equation}\label{step6}
\frac{3\bm{u}^{*} - 4\bm{u}^{n} + \bm{u}^{n-1}}{2\dt} = 2 \bm{F}^{n} - \bm{F}^{n-1}\\
\end{equation}

\emph{(iii)} Solve the Poisson equation for pressure  and correct the velocity field to satisfy incompressibility:
\begin{eqnarray}\label{step7}
\nabla^2 p^{n+1} & = & \frac{3}{2\dt}\nabla \cdot \bm{u}^*,\\
\label{step8} \bm{u}^{n+1} & = & \bm{u}^* - \frac{2}{3} \dt \nabla p^{n+1}.
\end{eqnarray}

\emph{(iv)} Solve the implicit equation for temperature:
\begin{equation}\label{step9}
\frac{3T^{n+1}-4T^n+T^{n-1}}{2\dt}=2G^n-G^{n-1}+\frac{1}{\Rey\Pran}\nabla\cdot\left(\kappa \nabla T^{n+1}\right).
\end{equation}

The impermeability boundary conditions for velocity are satisfied via the pressure correction as discussed below. The no-slip velocity boundary conditions at solid walls are imposed explicitly after (\ref{step8}). The boundary conditions for  potential and temperature at the walls are satisfied as a part of solution of the elliptic problems (\ref{step1}) and (\ref{step9}). 

For the computations conducted in this work, the structured grid is built on the lines of the cylindrical coordinate system (see figure \ref{fig:geom}) and clustered toward the walls and the interfaces between the layers according to the coordinate transformation successfully used in our recent work \cite{Krasnov:2012}. The clustering is a mixture of the Chebyshev-Gauss-Lobatto clustering and the uniform grid. It implements the advantages of a nearly-Chebyshev resolution  of boundary layers, while avoiding the strict time-step limitations caused by the smallest grid step near the boundary. 

In the radial direction, the coordinate transformation is 
 \begin{equation}\label{cheb_r}
    r=A_r\sin(\pi\eta/2)+(1-A_r)\eta, 
 \end{equation}
 where $0\le\eta\le 1$, $A_r=0.96$, and the grid is uniform in the $\eta$-coordinate. 
 
In the vertical direction, a similar transformation is applied separately to each layer in such a way that points are clustered towards both the upper and lower boundaries. Denoting the global transformed coordinate, in which the grid is uniform, as $-1\le\xi\le 1$ and introducing local coordinates 
 \begin{equation}\label{local_coor}
    \xi_{\ell}=2\frac{\xi-\xi_b}{\xi_t-\xi_b}-1,  \quad  z_{\ell}=2\frac{z-z_b}{z_t-z_b}-1,
 \end{equation}
 where $z_b$, $\xi_b$ and $z_t$, $\xi_t$ are the global coordinates of the bottom and top boundaries, we define the transformation as 
 \begin{equation}\label{cheb_z}
    z_{\ell}=A_z\sin(\pi\xi_{\ell}/2)+(1-A_z)\xi_{\ell}.
 \end{equation}
The blending coefficient is $A_z=0.96$  except for the case of thin electrolyte layer $H_E=0.1$, in which the grid is uniform, so $A_z=0$.

The total number of grid points in the vertical direction is divided equally among the three layers. 

Two major new features have been introduced into the algorithm for the purposes of this work. One of them concerns the velocity boundary condition, which now includes impermeability at not just solid walls, but also at the interfaces between the layers (see (\ref{intbc})). In order to enforce that we solve the pressure equation (\ref{step7}) separately in each of the three layers \textsf{A},   \textsf{E}, and  \textsf{B} and impose the Neumann condition obtained by projection of (\ref{step8}) on the normal to the boundary
\begin{equation}\label{neubc}
\frac{\partial p^{n+1}}{\partial n} =\frac{3}{2\Delta t}u^*_n
\end{equation} 
at the entire boundary of each domain, including the interfaces. The pressure fields obtained in this solution are defined up to additive constants, and can be combined into one continuous pressure field by adjusting the constants.

Another new feature concerns the solution of the elliptic problems (\ref{step1}) and (\ref{step9}) for electric potential and temperature. In the original and new versions of the algorithm, this is done using the Fast Fourier Transform in the azimuthal angle $\theta$ and the direct solution of the two-dimensional elliptic problems for the Fourier coefficients -- functions of $r$ and $z$. Implementation of this procedure for (\ref{step1}) and (\ref{step9}) applied to the entire flow domain requires modifications accounting for the presence of variable coefficients $\sigma(z)$ and $\kappa(z)$.   As discussed in detail in \cite{Krasnov:FD:2011}, the 2D elliptic problems for the Fourier coefficients are solved using the high-level routine of the FishPack library \cite{Fishpack} that discretizes a 2D separable elliptic equation on a uniform structured grid using central differences of the second order and solves the resulting matrix equation by the cyclic reduction method. In order to use this algorithm and retain the conservation properties of our scheme, we have to express the 2D elliptic equations in the transformed coordinates $\eta$-$\xi$ in such a way that the discretized equations produced by the FishPack routine would correspond exactly to the discretization of the original equations on the non-uniform grid on the $r$-$z$-plane. 

The procedure for the equations with constant coefficients is described in \cite{Krasnov:FD:2011}. For the case of variable coefficients, it is modified as illustrated by the following example. We divide the potential equation (\ref{step1}) by $\sigma(z)$ and require that  
the discretization on the non-uniform grid:
\begin{equation}
\left.\frac{1}{\sigma}\frac{\delta}{\delta z}\left(\sigma\frac{\delta \phi}{\delta z}\right)\right|_{z=z_j}=\frac{1}{\sigma_j}\frac{\sigma_{j+1/2}\frac{\phi_{j+1}-\phi_{j}}{z_{j+1}-z_{j}} -\sigma_{j-1/2}\frac{\phi_{j}-\phi_{j-1}}{z_{j}-z_{j-1}} }{z_{j+1/2}-z_{j-1/2}}
\end{equation}
is reproduced exactly by the central-difference approximation on the uniform grid of the same term expressed in the transformed coordinates (we use the same notation as in \cite{Krasnov:FD:2011}):
\begin{equation}
\left.\frac{1}{\sigma}\frac{\delta}{\delta z}\left(\sigma\frac{\delta \phi}{\delta z}\right)\right|_{z=z_j}=\frac{1}{c_3^3}\left(c_1\frac{\phi_{j+1}-2\phi_j+\phi_{j-1}}{\Delta \xi^2}-c_2 \frac{\phi_{j+1}-\phi_j-1}{2\Delta \xi}\right).
\end{equation}
In the formulas above, $\delta$ stands for the  central-difference operator, and the quantities at the half-integer points are obtained by interpolation,  e.g., as $z_{j+1/2}=(z_{j+1}+z_j)/2$. Direct comparison of the two formulas leads us to:
\begin{eqnarray}
c_1 & = & \frac{\sigma_{j+1/2}(z_j-z_{j-1})+\sigma_{j-1/2}(z_{j+1}-z_{j})}{2 \Delta \xi}\\
c_2 & = & \frac{\sigma_{j-1/2}(z_{j+1}-z_{j})+\sigma_{j+1/2}(z_{j}-z_{j-1})}{\Delta \xi^2}\\
c_3 & = &\sigma_j \frac{z_{j+1}-z_{j-1}}{2\Delta \xi}\frac{z_{j+1}-z_{j}}{\Delta \xi}\frac{z_{j}-z_{j-1}}{\Delta \xi}.
\end{eqnarray}

The other terms of the equations involving the variable coefficients $\sigma(z)$ and $\kappa(z)$ are discretized directly on the non-uniform grid in a straightforward manner.

The algorithm is parallelized using the OpenMP parallelization, with a typical simulation ran on 16 cpus of a shared-memory workstation.

A grid sensitivity study has been conducted for the system with $H=1.0$, $H_E=0.304$ and $z_E=0$. Comparing the exponential growth rates $\gamma$ obtained on various grids, we have determined that the grid with $N_r=60$, $N_z=60$, and $N_{\theta}=96$ is sufficient for accurate results. Further increase of the grid size by one third in each direction changes $\gamma$ by less than 2\%. 

A direct validation of the numerical model is impossible in our case, since no accurate and reliable experimental, numerical, or analytical data for the system considered in this paper are available. We can, however, rely on the extensive verification of the numerical method conducted in our earlier studies of magnetohydrodynamic shear and convection flows \cite{Krasnov:FD:2011,ZikanovJFM:2013,Zhang:2014,Lv:2014,Liu:2015,Krasnov:2012}. Furthermore, as demonstrated in section \ref{sec:disc}, the model's validity is supported by the consistency of its predictions with the known data for the convection instability in the Rayleigh-Benard layer and in the layer with internal heating. 

\section{Results}\label{sec:res}
The profile of the base-state temperature $T_0(z)$ illustrated in fig.~\ref{fig:geom}b suggests the possible mechanism of the instability. The temperature field is stratified unstably in the upper part of the electrolyte layer and, to a much smaller degree, in the layer \textsf{A}. If sufficiently strong, the stratification should cause convection flow, which may induce flow in the rest of the battery. In this section, we present the results of the computational analysis of this phenomenon. A discussion of the underlying physics and of the implications for the battery's operation are given, respectively, in sections \ref{sec:disc} and \ref{sec:impl}.
\label{sec:results}

\subsection{Detailed analysis of a convection flow}\label{sec:typical}
\begin{figure}
\begin{center}
\includegraphics[width=0.5\textwidth]{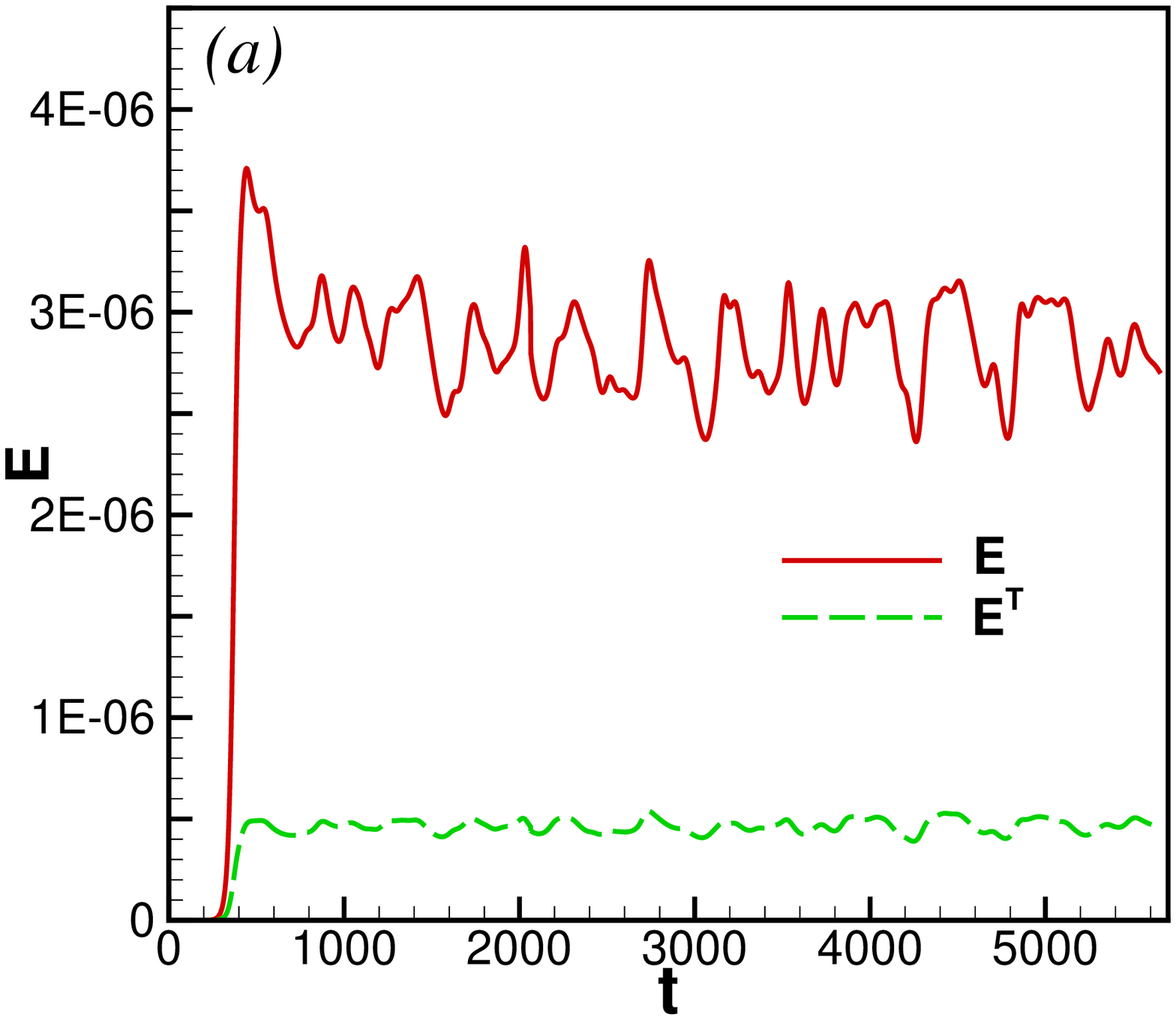}\includegraphics[width=0.5\textwidth]{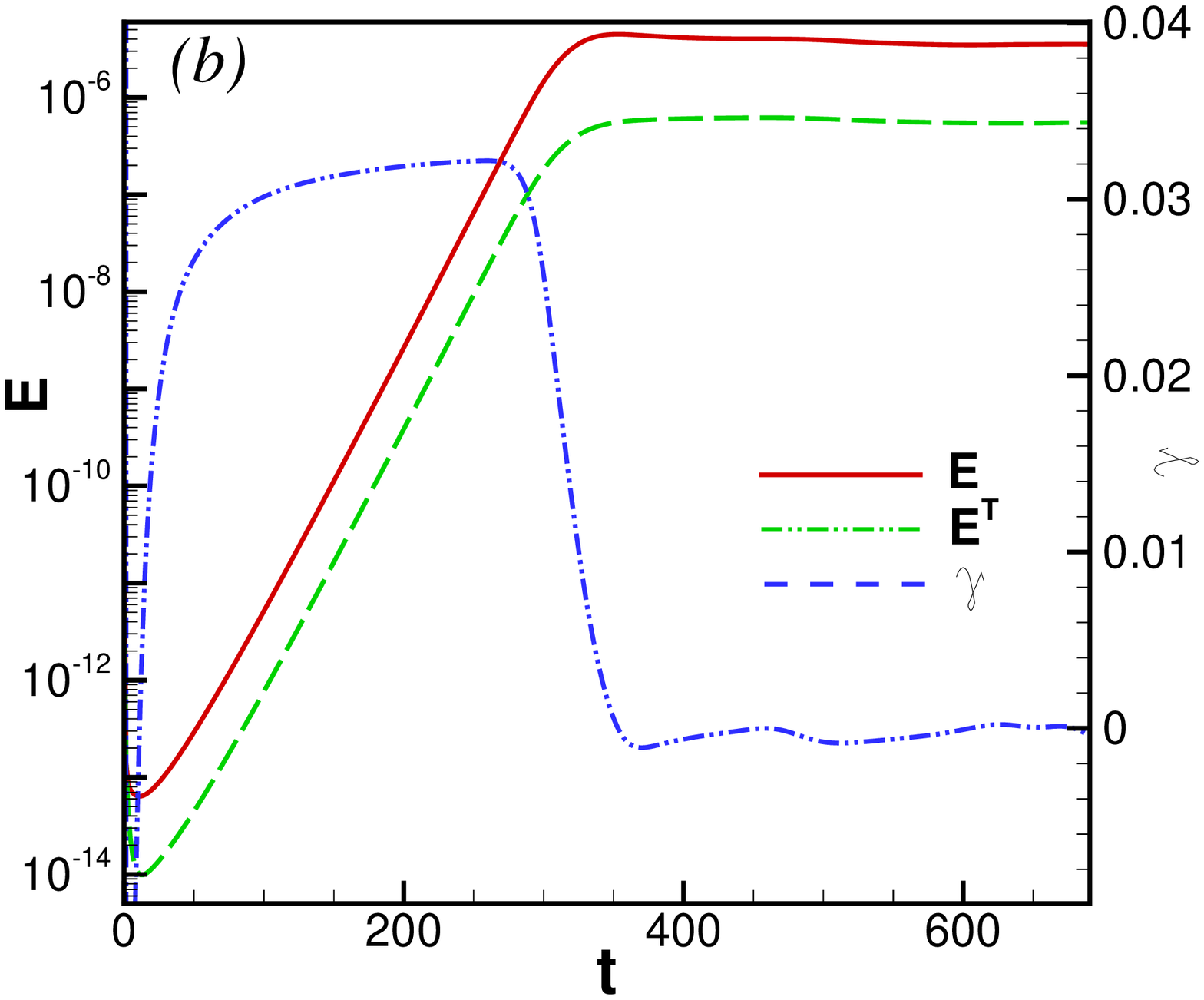}
\caption{\emph{(a)}, Volume-averaged kinetic and thermal energies of perturbations (see (\ref{perten})) in the simulation with $H_E=0.304$ and $\Gras=4.5\times 10^6$. 
\emph{(b)},  Perturbation energies and the growth rate $\gamma$ (see (\ref{gamma})) during and shortly after the phase of exponential growth. }
\label{fig:perten}
\end{center}
\end{figure}

As a typical example of the instability and the resulting convection flow, we consider the case $H_E=0.304$ and $\Gras=4.5\times 10^6$. The evolution of the volume-averaged perturbation energies 
(\ref{perten}) is shown in Fig.~\ref{fig:perten}. We see that after a short initial period the perturbations become dominated by the strongest instability mode, and the energies grow exponentially. The growth continues till about $t=350$, when the levels $E\sim 3\times 10^{-6}$, $E_{T}\sim 5\times 10^{-7}$, corresponding, in our units, to the finite-amplitude saturation of the instability, are reached. After that, the flow's evolution is nonlinear, with slow and irregular oscillations of energies around constant values.

\begin{figure}
\begin{center}
\includegraphics[width=0.5\textwidth]{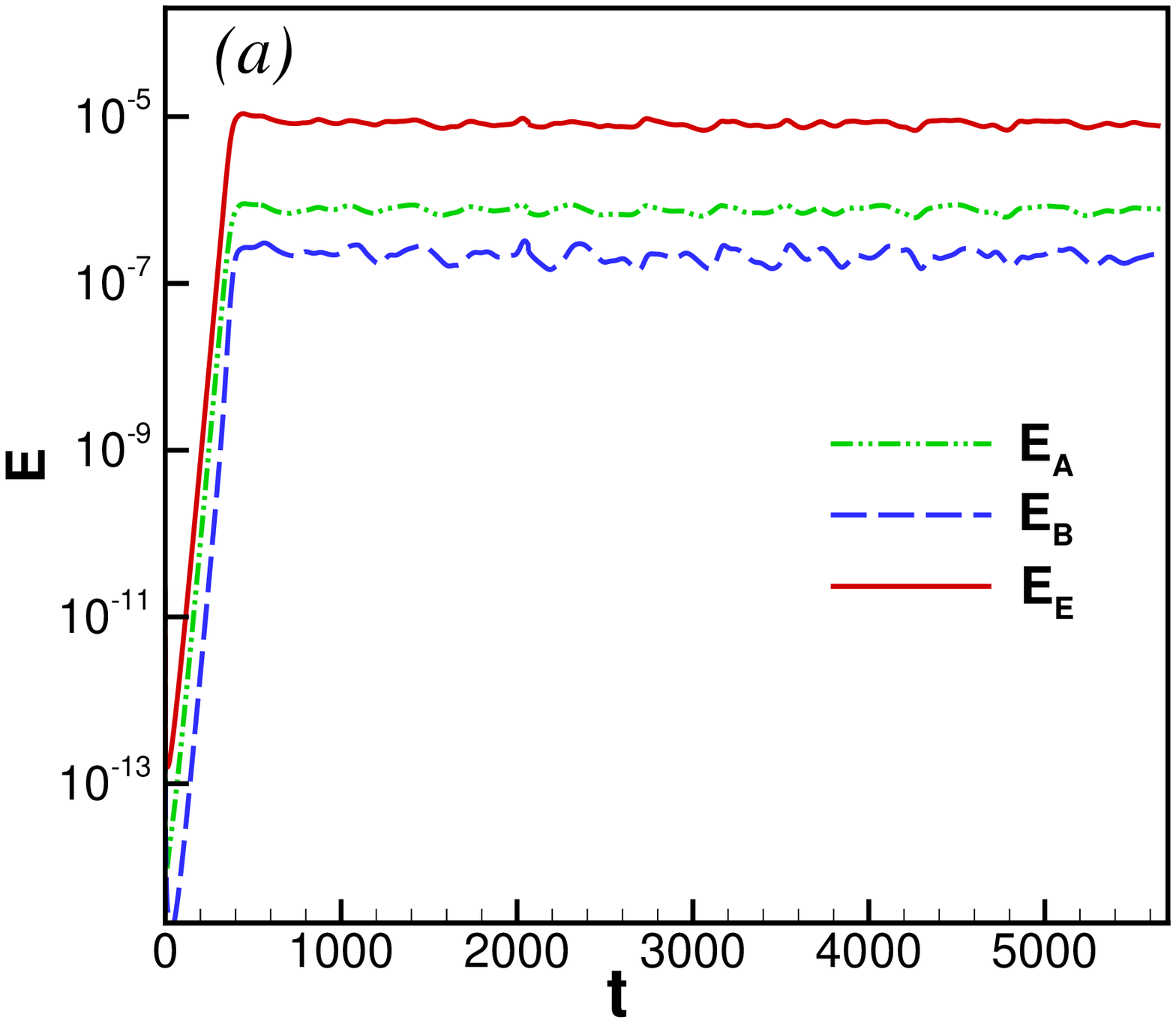}\includegraphics[width=0.5\textwidth]{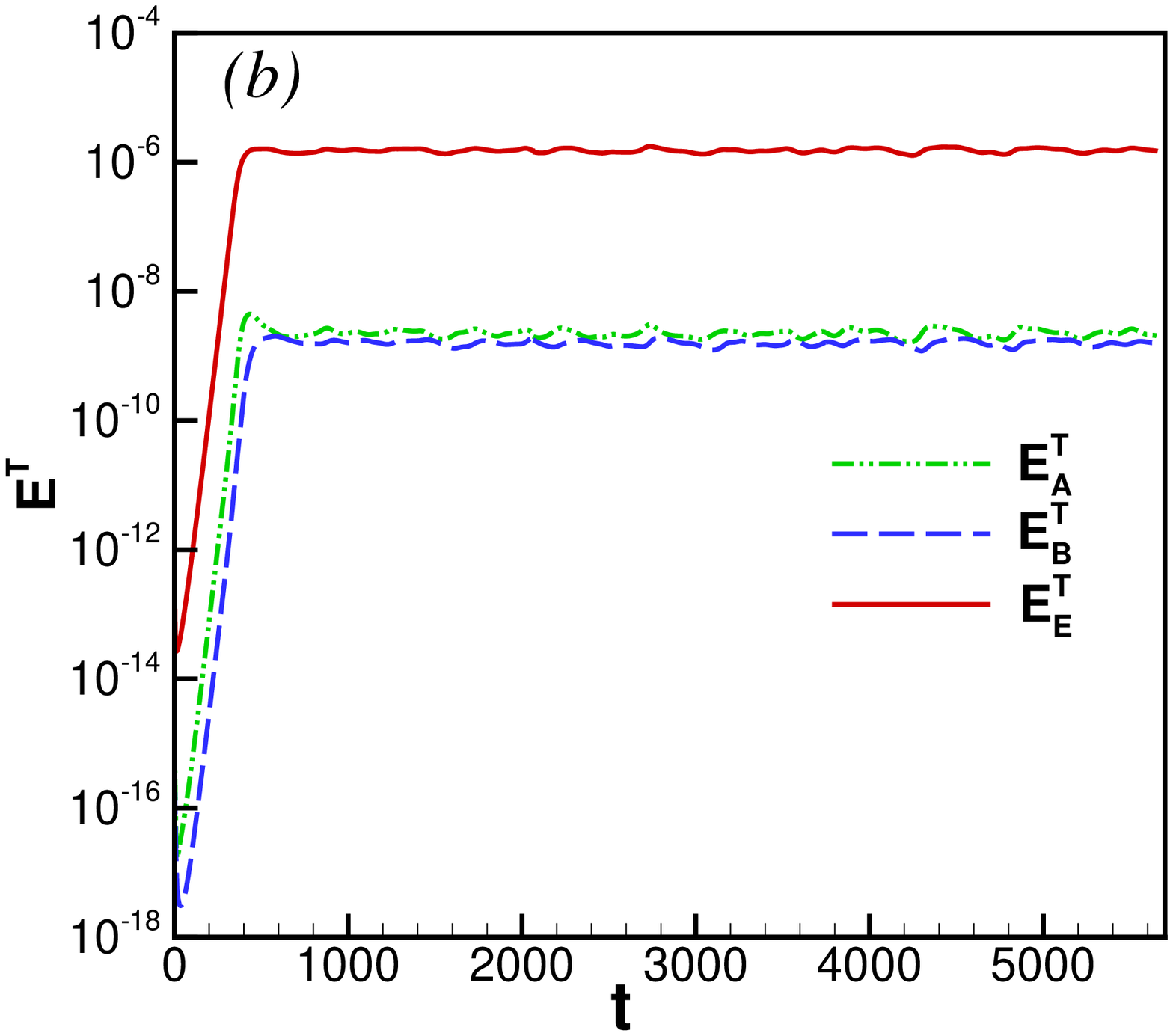}
\caption{Kinetic \emph{(a)} and thermal \emph{(b)} energies of perturbations averaged over individual layers in the simulation with $H_E=0.304$ and $\Gras=4.5\times 10^6$. 
}
\label{fig:pertenlayer}
\end{center}
\end{figure}

The perturbation energies computed as in (\ref{perten}), but with the averaging performed over separated layers \textsf{A}, \textsf{B}, and \textsf{E}, are shown in Fig.~\ref{fig:pertenlayer}. We see that the amplitude of the perturbations is the highest in the electrolyte and the lowest in the bottom layer \textsf{B}. Specifically, in the developed flow, we find, for the time-averaged values, $E_A/E_E\approx 0.096$, $E_B/E_E\approx 0.024$.

\begin{figure}
\begin{center}
\includegraphics[width=0.5\textwidth]{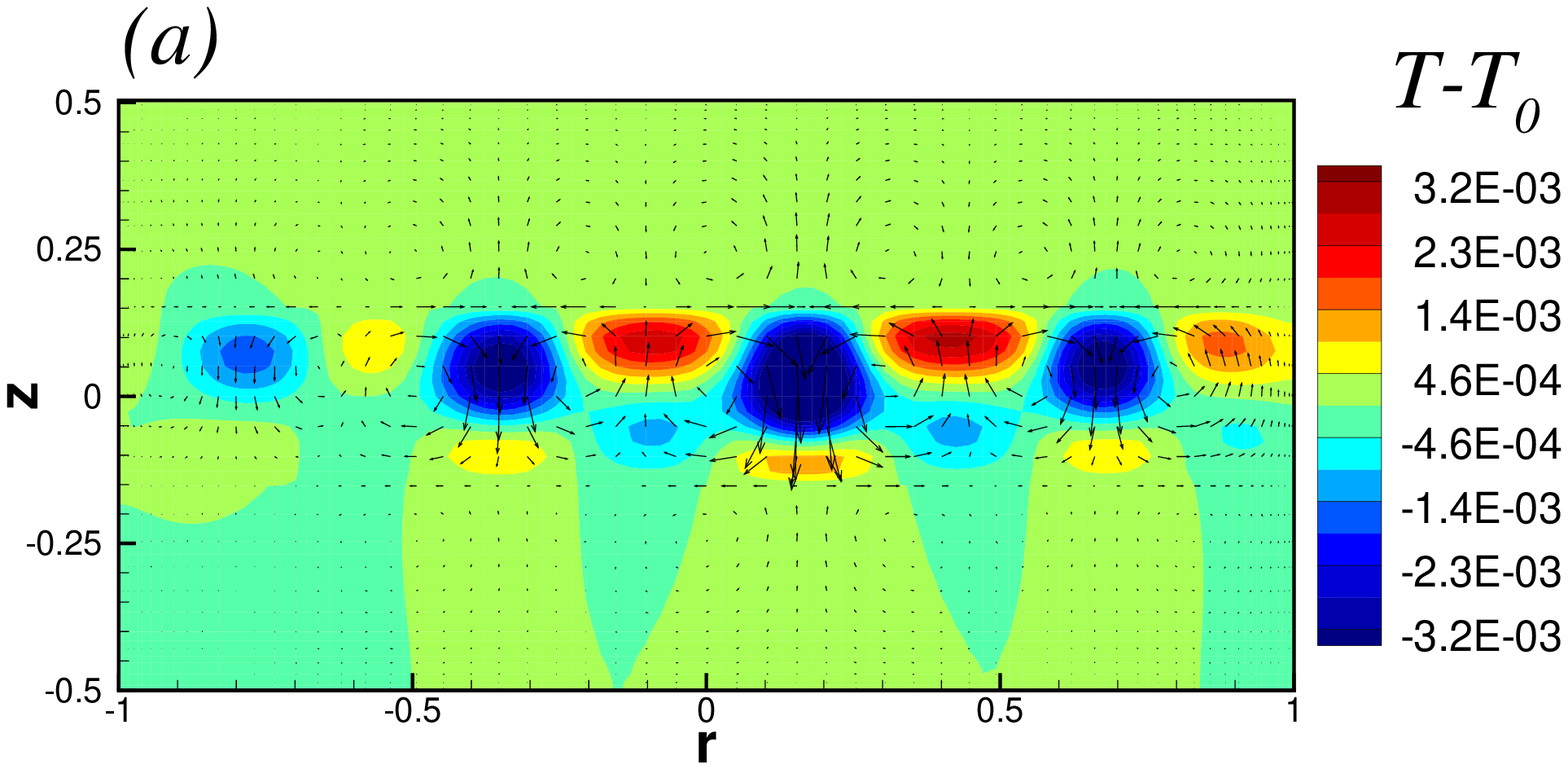}\includegraphics[width=0.5\textwidth]{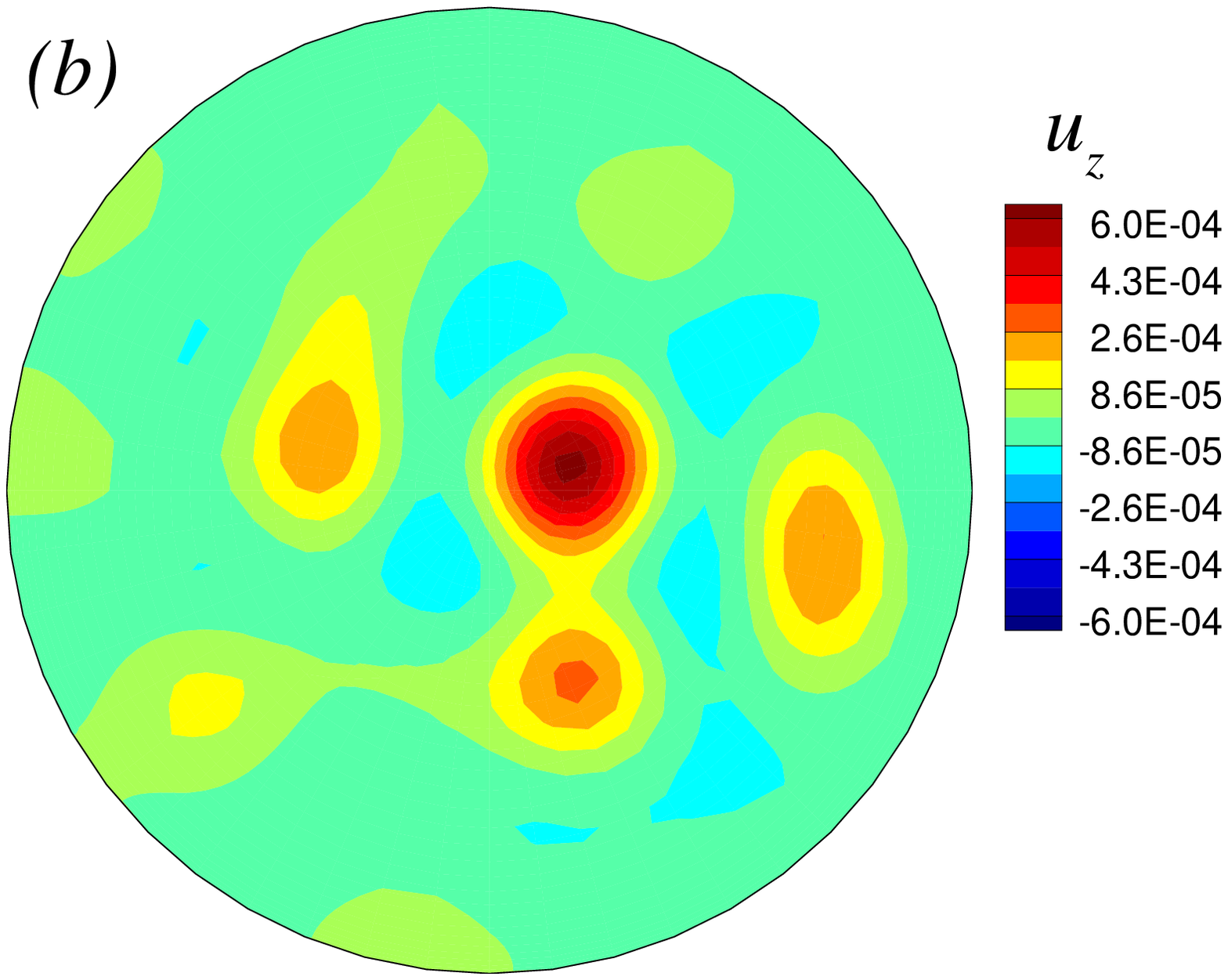}\\
\includegraphics[width=0.5\textwidth]{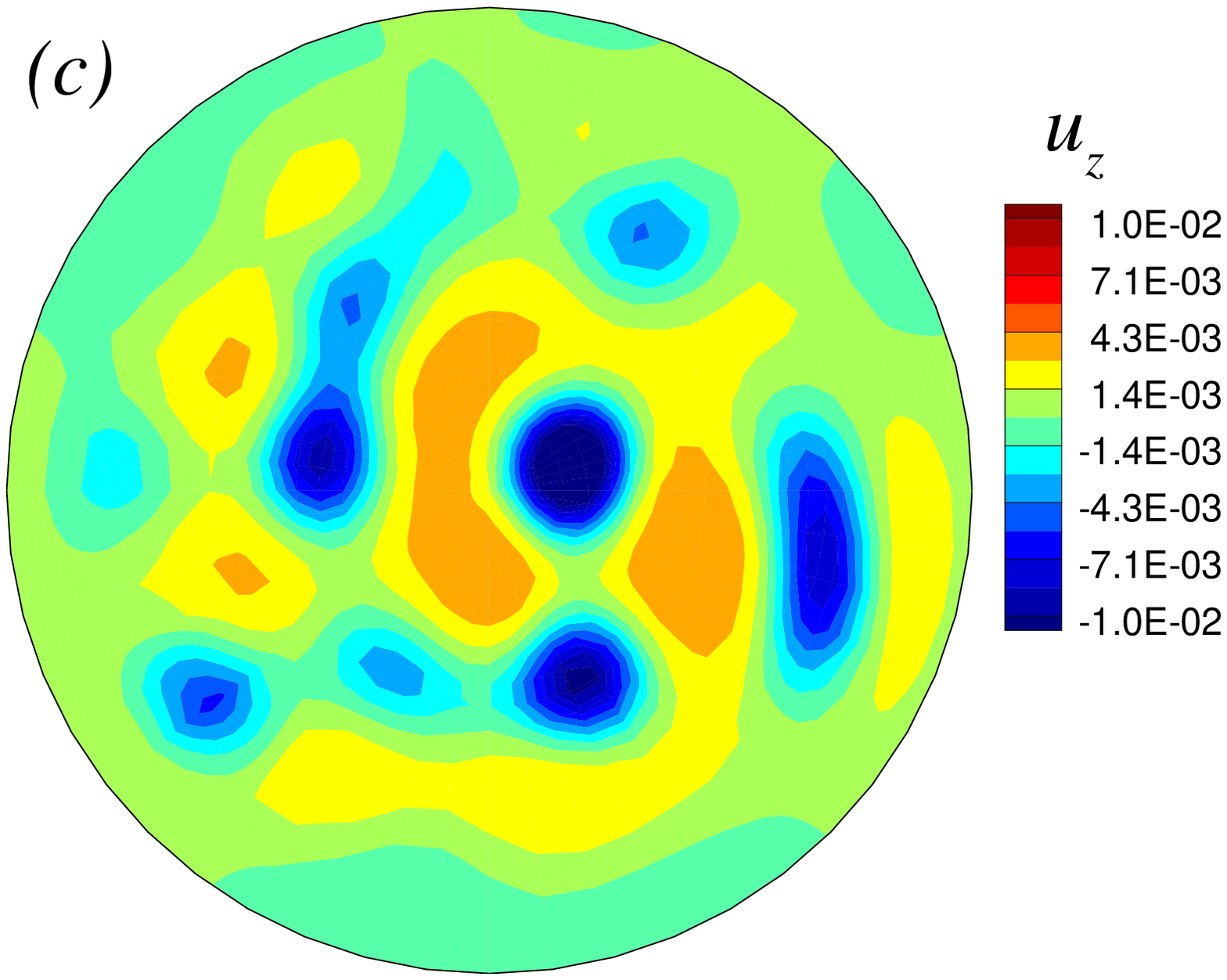}\includegraphics[width=0.5\textwidth]{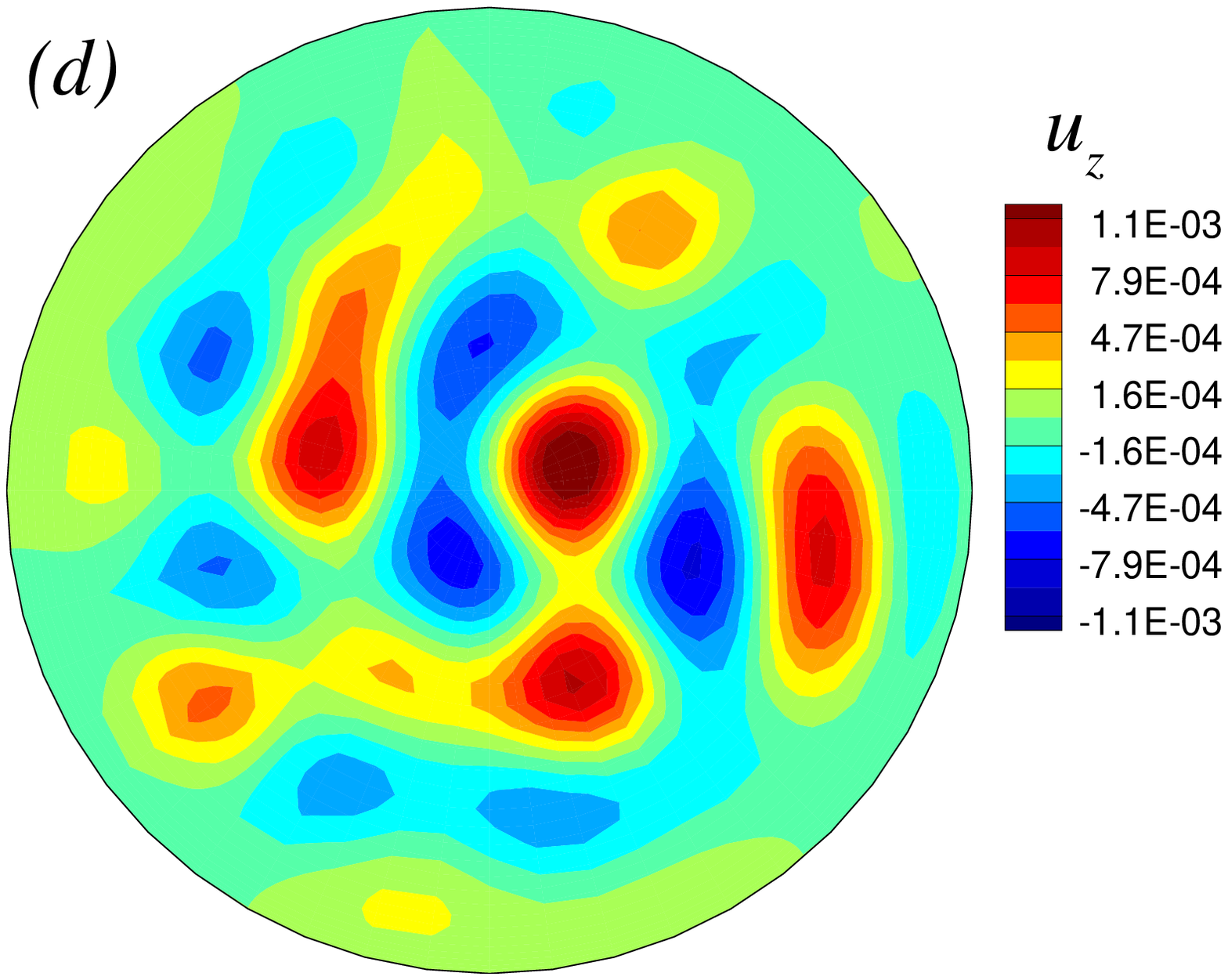}
\caption{Spatial structure of perturbations during the phase of exponential growth obtained in the simulation with $H_E=0.304$ and $\Gras=4.5\times 10^6$ at $t=300$. \emph{(a)}, Temperature perturbations $T-T_0$ and velocity vectors (drawn at every second grid point in each direction) in the vertical cross-section $\theta=0$, $\pi$. \emph{(b)}--\emph{(d)}, vertical velocity in the horizontal cross-sections $z=0.25$  \emph{(b)}, $z=0$ \emph{(c)}, and $z=0.75$ \emph{(d)}.
}
\label{fig:sp1}
\end{center}
\end{figure}

\begin{figure}
\begin{center}
\includegraphics[width=0.5\textwidth]{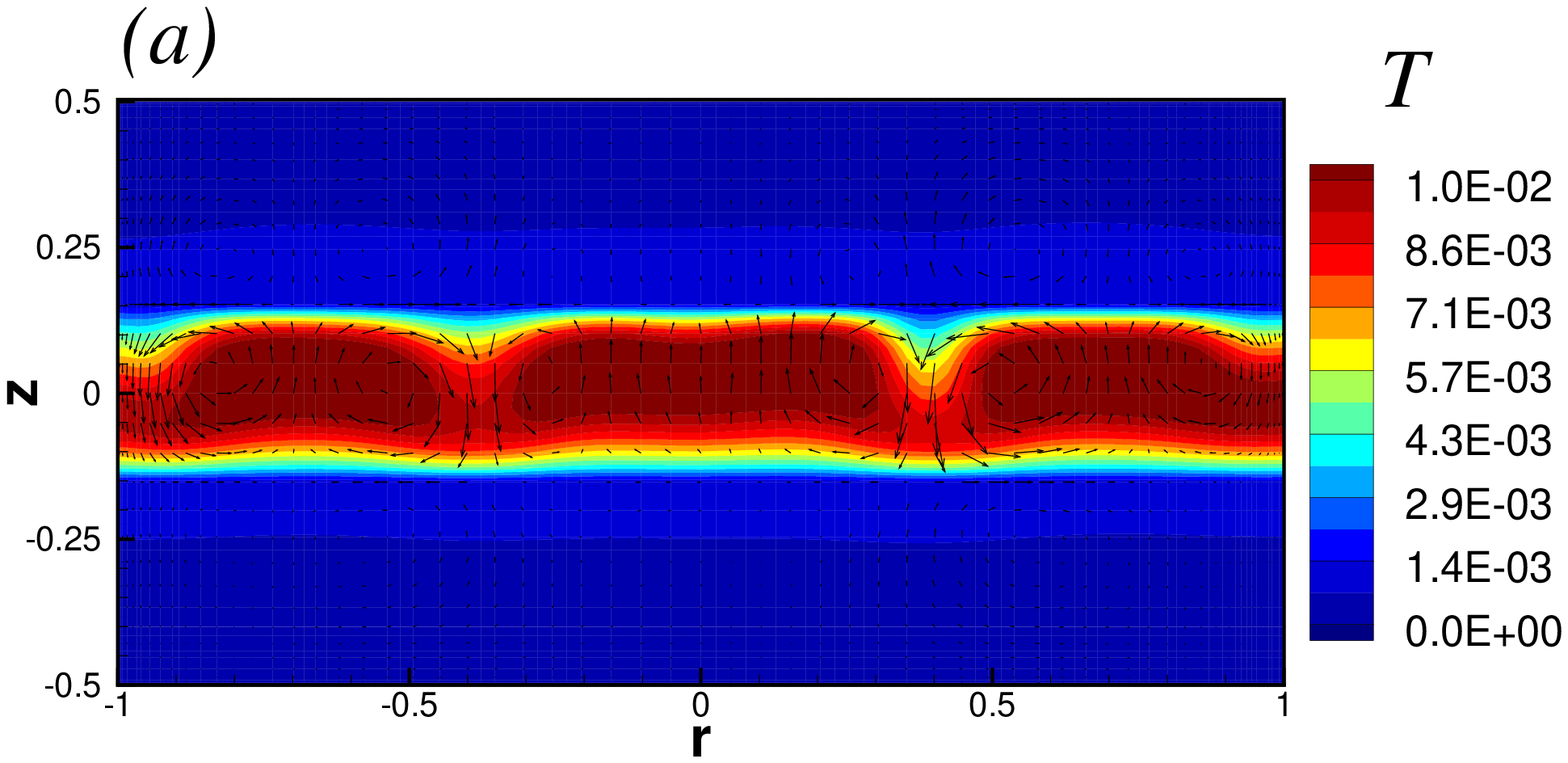}\includegraphics[width=0.5\textwidth]{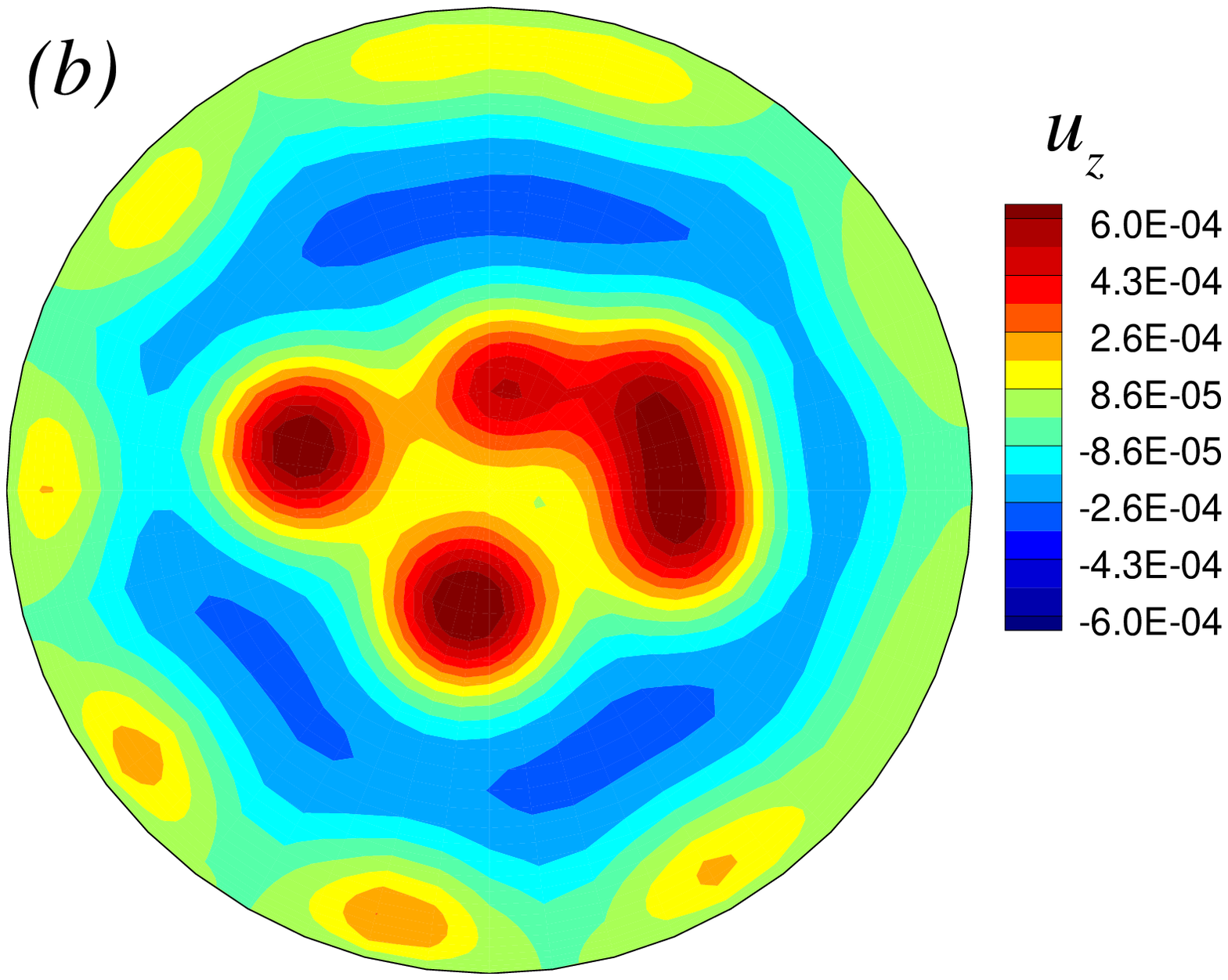}\\
\includegraphics[width=0.5\textwidth]{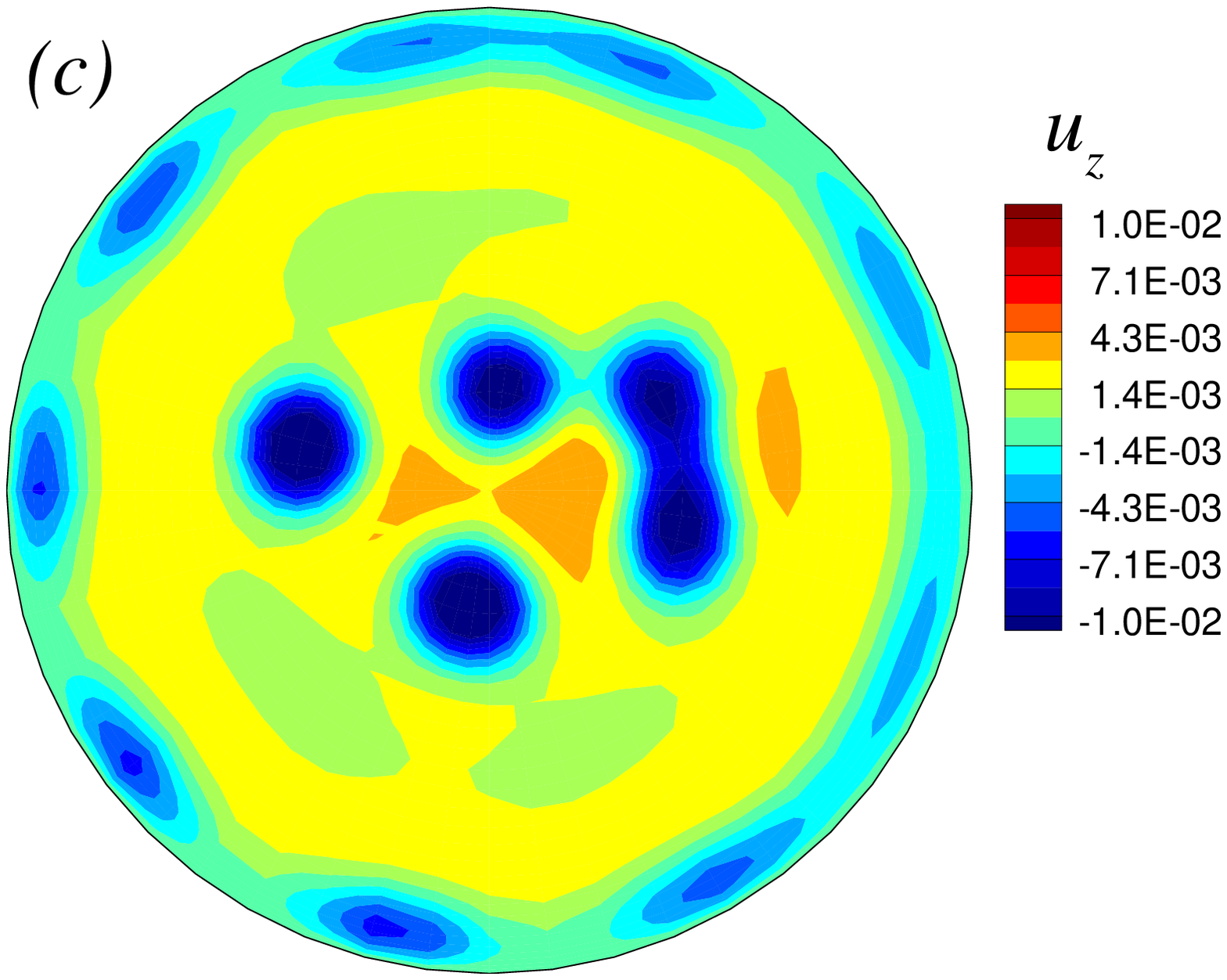}\includegraphics[width=0.5\textwidth]{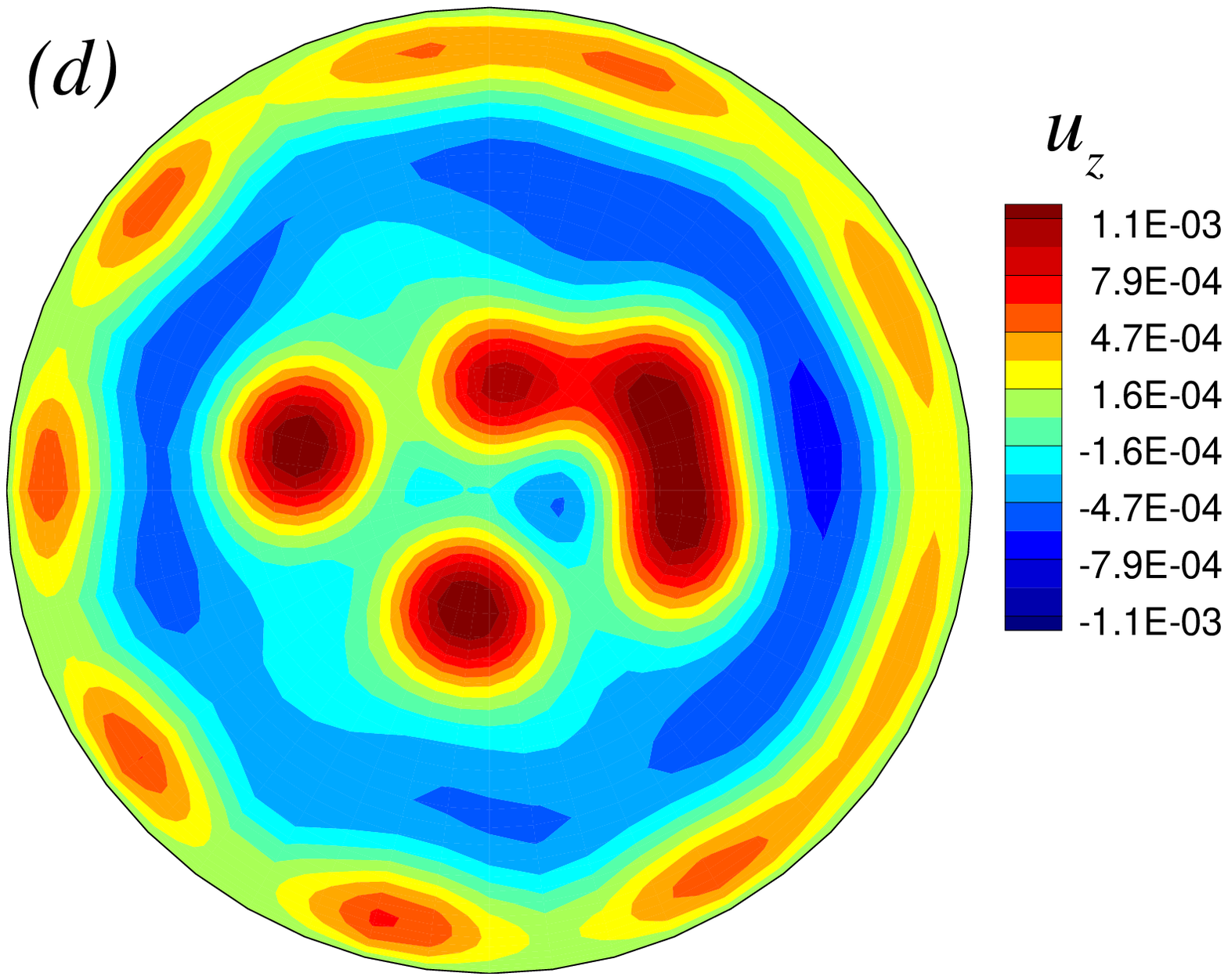}
\caption{Spatial structure of developed finite-amplitude convection flow obtained in the simulation with $H_E=0.304$ and $\Gras=4.5\times 10^6$ at $t=1500$. \emph{(a)}, Full temperature  $T$ and velocity vectors (drawn at every second grid point in each direction) in the vertical cross-section $\theta=0$, $\pi$. \emph{(b)}--\emph{(d)}, vertical velocity in the horizontal cross-sections $z=0.25$  \emph{(b)}, $z=0$ \emph{(c)}, and $z=0.75$ \emph{(d)}.
}
\label{fig:sp2}
\end{center}
\end{figure}

In order to illustrate the spatial structure of the flow, Figs.~\ref{fig:sp1} and \ref{fig:sp2} show distributions of temperature and velocity in the vertical (drawn through the axis) and horizontal cross-sections. 

The exponentially growing dominant mode of the linear instability (see Fig.~\ref{fig:sp1}) is a combination of several three-dimensional convection cells located mainly within the electrolyte layer. Much weaker motion is generated in the top layer \textsf{A}. The motion in the bottom layer \textsf{B} is even weaker. The convection cells do not form any regular pattern in the horizontal plane and have the typical horizontal size about an order of magnitude smaller than the diameter of the battery.

At the nonlinear saturation stage, the convection cells gradually evolve into the pattern illustrated in Fig.~\ref{fig:sp2}. Individual unsteady cells are still observed, but the flow acquires a degree of axial symmetry. In the electrolyte layer, the zones of downward flow are concentrated near the sidewalls and around the axis, while the upward flow primarily occurs in a ring at middle values of $r$. The weaker circulation in the bottom and top layers has a similar structure, but the opposite circulation sign.

To evaluate the effect of convection flow and the mixing associated with it on the heat transfer in a battery, we compute the two Nusselt numbers corresponding to the heat transfer through the top and the bottom walls:
\begin{equation}
\label{nusselt}
\Nu_{A}=\frac{\left.\partial \langle T\rangle/\partial z\right|_{z=H}}{\left.\partial  T_0/\partial z\right|_{z=H}}, \quad \Nu_{B}=\frac{\left.\partial \langle T\rangle/\partial z\right|_{z=0}}{\left.\partial  T_0/\partial z\right|_{z=0}},
\end{equation}
where $\langle T \rangle$ is the instantaneous temperature field averaged over the entire wall surface

In a steady state, the total heat flux through both walls remains equal to the rate of internal heat generation $Q_0$. This implies that the cumulative Nusselt number
\begin{equation}
\label{nusselt_cum}
\Nu=\frac{\left.\partial \langle T\rangle/\partial z\right|_{z=H}+\left.\partial \langle T\rangle/\partial z\right|_{z=0}}{\left.\partial  T_0/\partial z\right|_{z=H}+\left.\partial  T_0/\partial z\right|_{z=0}}=\frac{1}{2}\left(\Nu_A+\Nu_B \right)
\end{equation}
must be equal to one. 

\begin{figure}
\begin{center}
\includegraphics[width=0.5\textwidth]{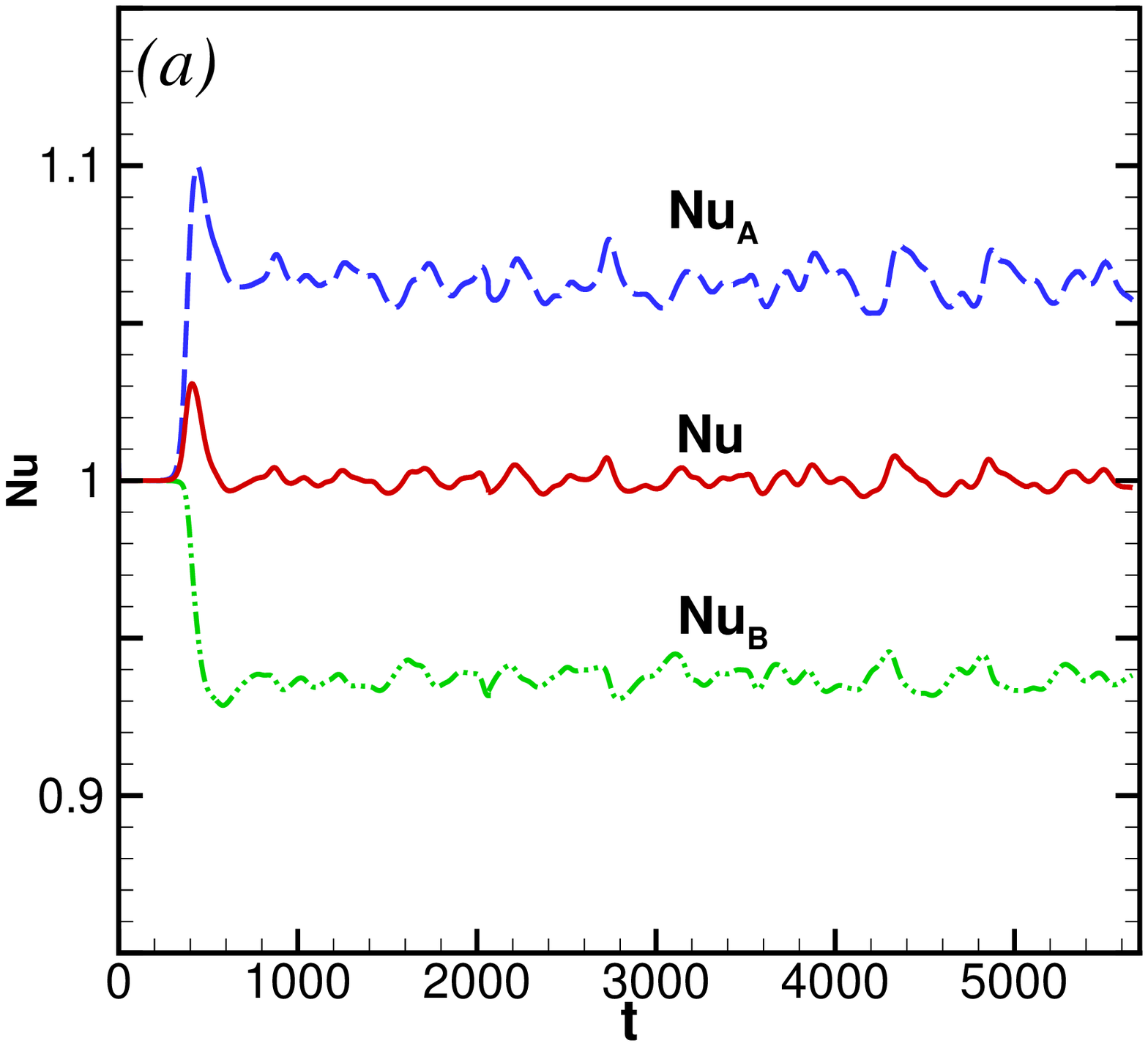}\includegraphics[width=0.5\textwidth]{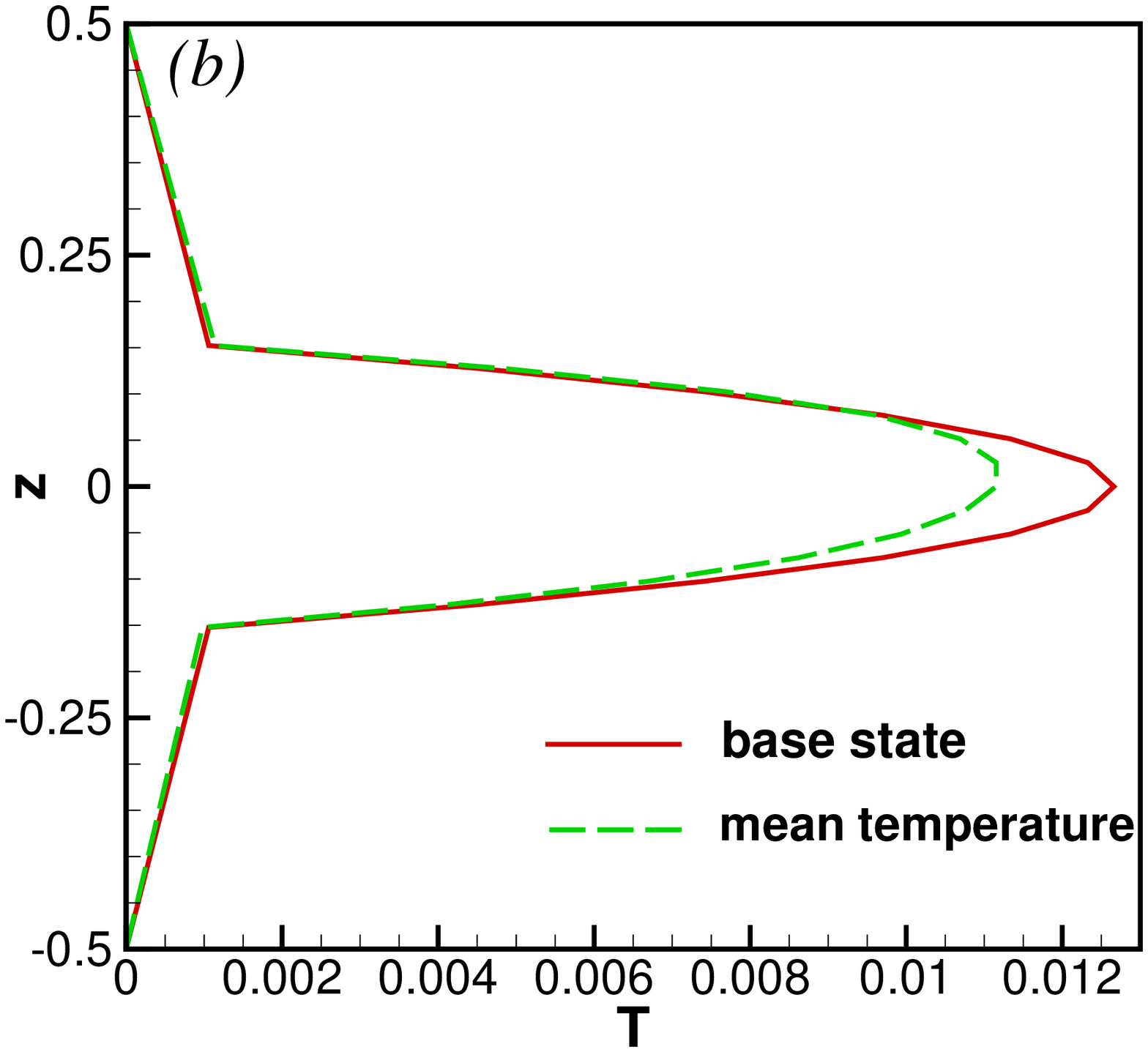}
\caption{\emph{(a)}, The Nusselt numbers at the top $\Nu_A$ and bottom $\Nu_B$ walls (see (\ref{nusselt})) and the cumulative Nusselt number $\Nu$ (see (\ref{nusselt_cum})) computed in the simulation with $H_E=0.304$ and $\Gras=4.5\times 10^6$. \emph{(b)}, Distribution of mean temperature $\overline{T}$ at the stage of fully developed flow for the same parameters. The temperature profile in the base state $T_0(z)$ is shown for comparison.}
\label{fig:nusselt}
\end{center}
\end{figure}

The evolution of $\Nu_A$ and $\Nu_B$ with time for our typical case is shown in Fig.~\ref{fig:nusselt}a. We see that, when a finite-amplitude convection flow develops, the heat transfer through the top wall exceeds the conduction heat transfer by about 8\%, while the heat transfer through the bottom wall is reduced by a similar margin. The cumulative Nusselt number $\Nu$ fluctuates around unity confirming that our system's behavior, while unsteady, corresponds to a statistically steady state. 

The asymmetry of the heat flux in the presence of convection can be explained by the mixing effect of the convection flow. Fig.~\ref{fig:nusselt}b shows the distribution of the mean temperature $\overline{T}(z)$, which we  obtain by horizontal averaging and time averaging over the entire period of the evolution of fully developed flow. We see that the change  is such that the amplitude of $\partial\overline{T}/\partial z$ is slightly increased at the top wall and slightly decreased at the bottom wall.

\subsection{Parametric study}\label{sec:parametric}
As discussed in section \ref{sec:parameters}, we analyze how the convection instability changes with the two parameters of the system: the Grashof number $\Gras$ and the non-dimensional thickness of electrolyte $H_E$.  The aspect ratio of the battery is kept at $H=1$, the electrolyte layer is located in the middle of the cell, and the Hartmann number is related to $\Gras$ by (\ref{hagra}). In every simulation, the evolution of the flow is simulated and analyzed in the same manner as for the typical example presented in the previous section. We have found that in every case, except whose where the base state is stable, the flow evolves through the same stages of exponential growth and nonlinear saturation. 

The results are summarized in Table \ref{table1}, which lists all the completed simulations and shows the computed quantitative properties: the exponential growth rate $\gamma$ (see (\ref{gamma})), and the time-averaged perturbations energies $E$, $E^T$ (see (\ref{perten})), $E_A$, $E_B$, $E_E$, and the Nusselt numbers $\Nu_{A}$, $\Nu_{B}$ (see (\ref{nusselt})) obtained for fully-developed flows. 

\begin{table}
  \centering 
\begin{tabular}{cc|cccccccc}
$H_E$ & $\Gras$ & $\gamma$ & $E$ & $E^T$ & $E_A$ & $E_B$ & $E_E$ & $\Nu_A$ & $\Nu_B$ \\
\hline
0.1 & 3.0E8 & stable &  & & & & & & \\
0.1 & 5.0E8 & 0.0005&           &             &           &           &          &       &     \\
0.1 & 7.0E8 & 0.0022& 2.98E-6& 1.23E-9 & 6.49E-6 & 3.32E-8 & 4.88E-7 & 1.02 & 0.98 \\
0.1 & 1.0E9 & 0.008 & 7.26E-6 & 4.65E-9 & 1.57E-5 & 8.28E-8 & 1.47E-6 & 1.07 & 0.93 \\
0.1 & 2.0E9 & 0.024 & 1.29E-5 & 1.29E-8 & 2.78E-5 & 1.39E-7 & 2.80E-6 & 1.16 & 0.85\\
0.1 & 3.0E9 & 0.033 & 1.28E-5 & 1.59E-8 & 2.76E-5 & 1.37E-7 & 3.01E-6 & 1.19 & 0.82 \\
\hline
 0.304 & 2.0E6 & stable &   &   &   &  &  &  &  \\
 0.304 & 2.5E6 & 0.004 & 9.38E-7 & 1.27E-7 & 2.49E-7 & 5.83E-8 & 2.73E-6 & 1.02 & 0.98 \\
 0.304 & 3.0E6 & 0.013 & 1.81E-6 & 2.26E-7 & 4.68E-7 & 1.51E-7 & 5.23E-6 & 1.04 & 0.96 \\
 0.304 & 3.5E6 & 0.020 & 2.30E-6 & 3.32E-7 & 6.10E-7 & 1.82E-7 & 6.66E-6 & 1.05 & 0.95 \\
 0.304 & 4.0E6 & 0.027 & 2.80E-6 & 4.63E-7 & 7.63E-7 & 2.16E-7 & 8.11E-6 & 1.06 & 0.94 \\
 0.304 & 4.5E6 & 0.032 & 2.85E-6 & 4.65E-7 & 7.60E-7 & 2.17E-7 & 8.23E-6 & 1.06 & 0.94 \\
  \hline
 0.58 & 2.0E5 & stable & & & & & & & \\
 0.58 & 2.4E5 & 0.0016 & & & & & & & \\
  0.58 & 2.6E5 & 0.007 & & & & & & & \\
 0.58 & 2.8E5 & 0.052 & 3.25E-5 & 1.76E-5 & 2.23E-6 & 4.63E-7 & 5.78E-5 & 1.09 & 0.91\\
 0.58 & 3.5E5 & 0.07 & 4.57E-5 & 2.10E-5 & 4.89E-6 & 6.68E-7 & 7.67E-5 & 1.12 & 0.89\\
 0.58 & 5.0E5 & 0.09 & 4.75E-5 & 3.77E-5 & 6.16E-6 & 3.46E-7 & 7.96E-5 & 1.16 & 0.84\\
\end{tabular}
  \caption{Summary of the parametric study. The rate of exponential growth $\gamma$ (see (\ref{gamma})), and the time-averaged perturbations energies $E$, $E^T$ (see (\ref{perten})), time-averaged perturbation kinetic energies $E_A$, $E_B$, $E_E$,  and the Nusselt numbers $\Nu_{A}$, $\Nu_{B}$ (see (\ref{nusselt})) of fully-developed flows are shown for all the completed simulations. Slight deviations from the integral relation between $E_A$, $E_B$, $E_E$, and $E$ and from the relation $\Nu=1$ (see (\ref{nusselt_cum})) are due to the time averaging error. In several cases with weak instability (small values of $\gamma$) extension of the simulations beyond the exponential growth stage required very long runs and was not conducted.}\label{table1}
\end{table}

The critical Grashof numbers $\Gras_{cr}$, such that the base state is stable at $\Gras<\Gras_{cr}$ and unstable at $\Gras>\Gras_{cr}$, can be approximately determined by extrapolating the function $\gamma(\Gras)$ to $\gamma=0$ and verifying that the base state is stable at smaller $\Gras$. The results are presented in Table \ref{table2}. We see that, in term of our parameters, the batteries become more stable (requiring larger $\Gras$ for the instability) as the non-dimensional thickness of the electrolyte becomes smaller. This can be explained by the fact that the Grashof number is defined using the battery's radius $R$ as the length scale (see (\ref{grashof})). The approach is justified, because it establishes a direct link between the convection instability and the size and total current of the battery. At the same time, it is not consistent with the nature of the instability, which is caused by vertical stratification of the base temperature $T_0(z)$ in the electrolyte layer and, therefore, has $H_E$ as the relevant length scale. To address that, the table \ref{table2} also shows the critical values of the $H_E$-based Grashof number
\begin{equation}
\Gras^E=\Gras H_E^5.
\label{grasE}
\end{equation}
We see that these values are closer to each other. An increase of $\Gras^E_{cr}$  with growing $H_E$ can be attributed to the increasing constraining effect of the sidewalls. This aspect of our results is further discussed in section \ref{sec:disc}.

\begin{table}
  \centering 
\begin{tabular}{c|ccc}
$H_E$ & 0.1 & 0.304 & 0.58  \\
\hline
$\Gras_{cr}$ & 4.67E8E8 & 2.28E6 &  2.30E5 \\
$\Gras^E_{cr}$ & 4.67E3E3 & 5.91E3 &   1.51E4\\
\end{tabular}
  \caption{Instability thresholds  $\Gras_{cr}$ and   $\Gras^E_{cr}$ (see (\ref{grasE})) obtained by extrapolation of the curves $\gamma(\Gras)$ from table \ref{table1}.}\label{table2}
\end{table}

\begin{figure}
\begin{center}
\includegraphics[width=0.5\textwidth]{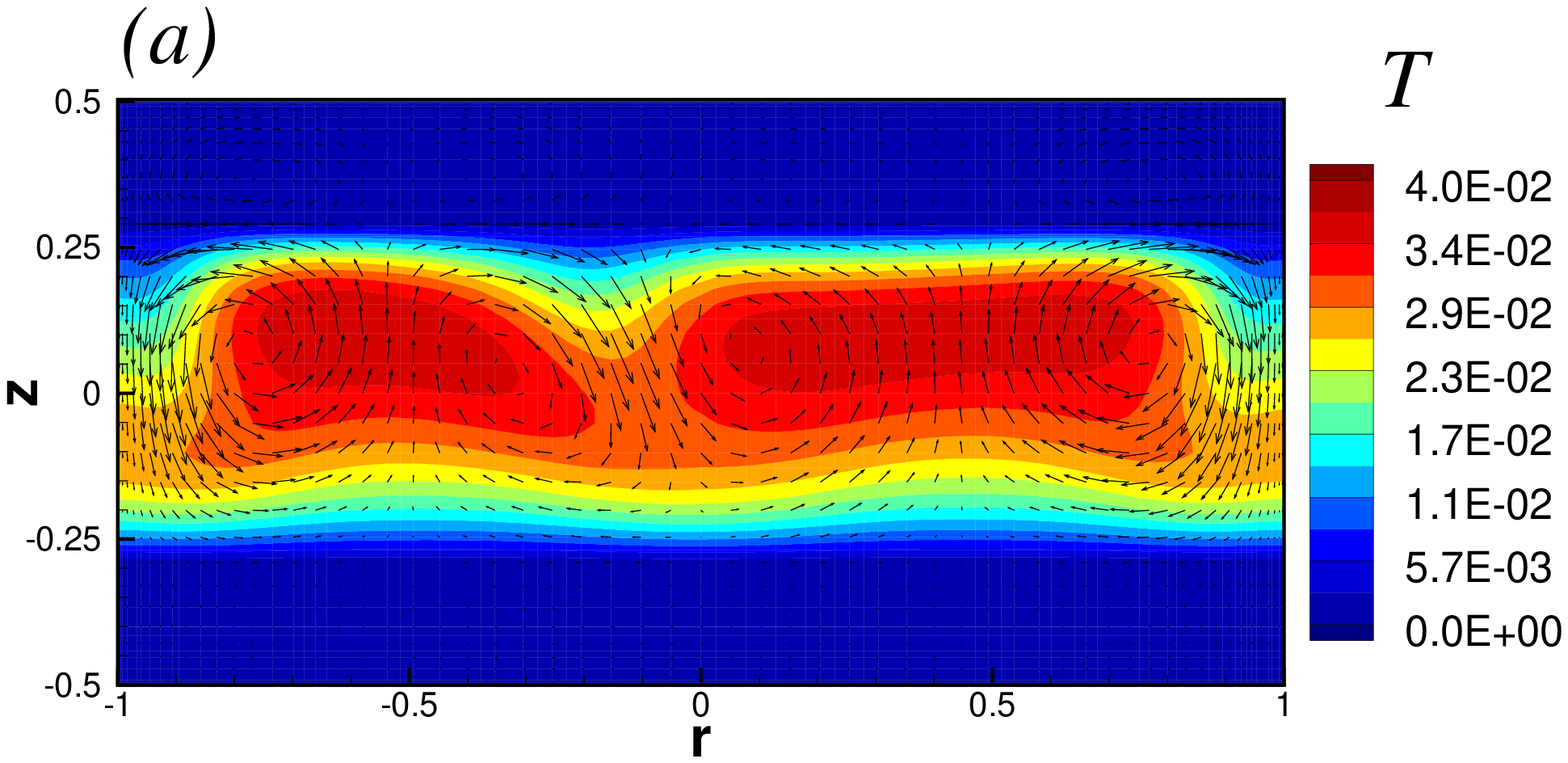}\includegraphics[width=0.5\textwidth]{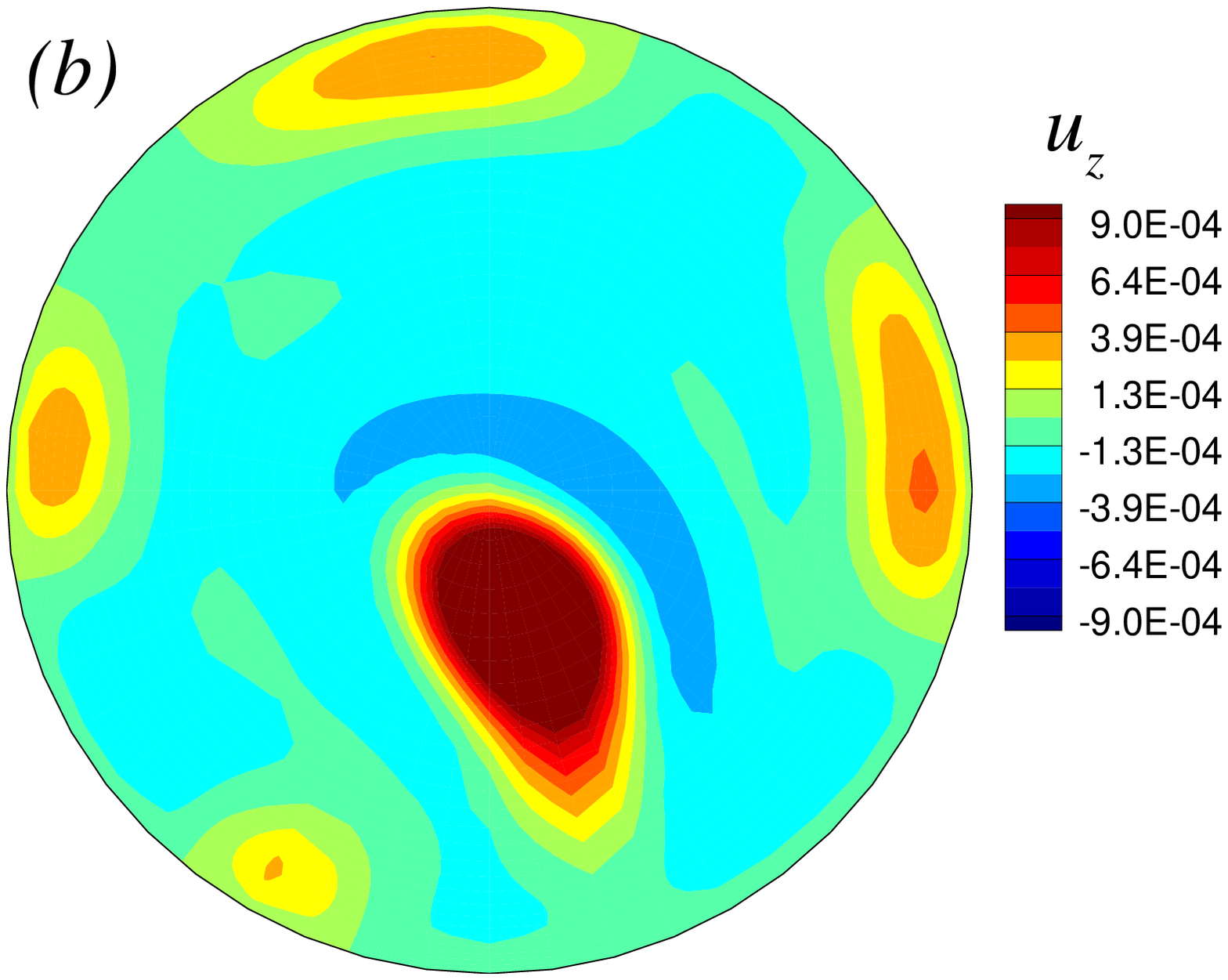}\\
\includegraphics[width=0.5\textwidth]{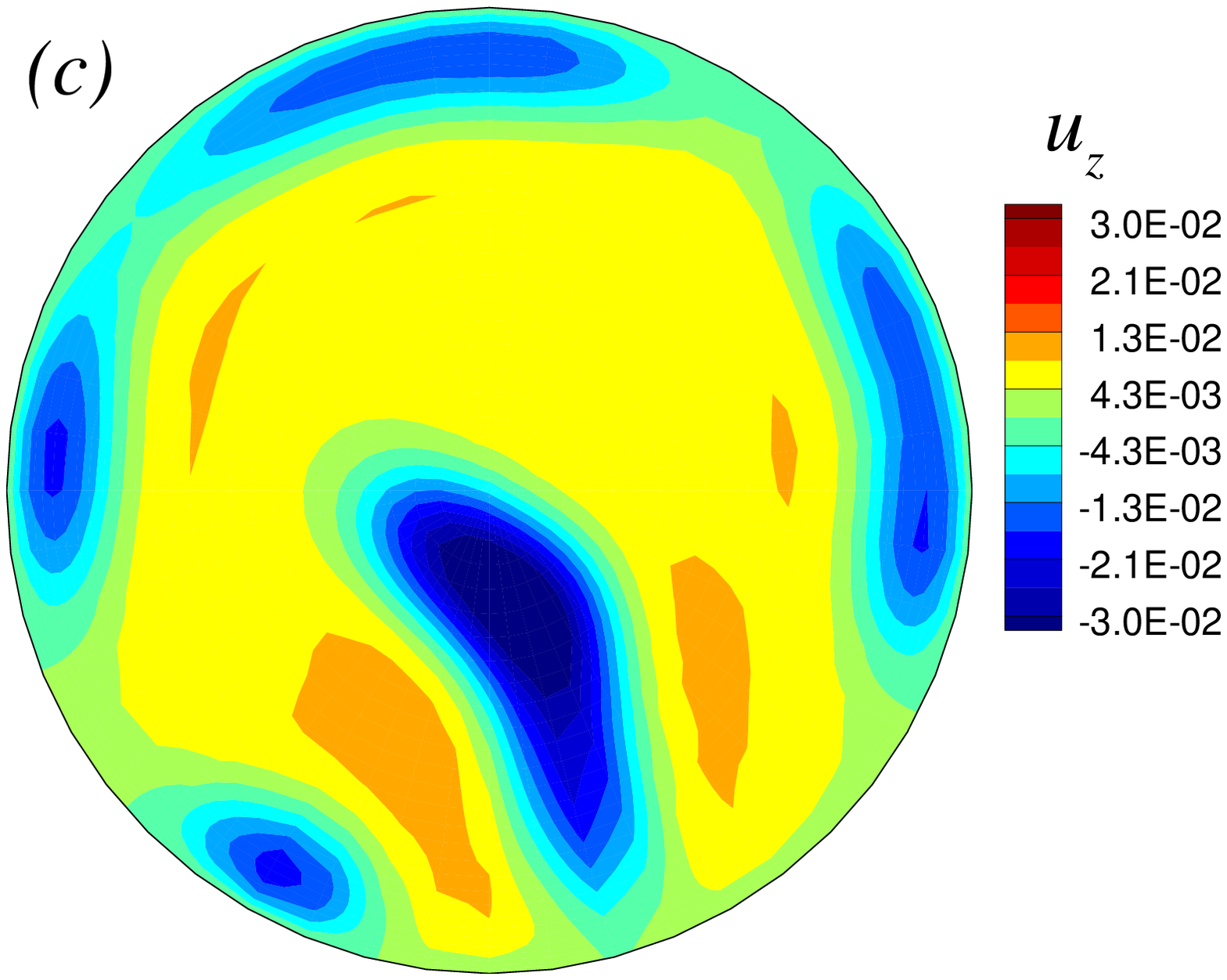}\includegraphics[width=0.5\textwidth]{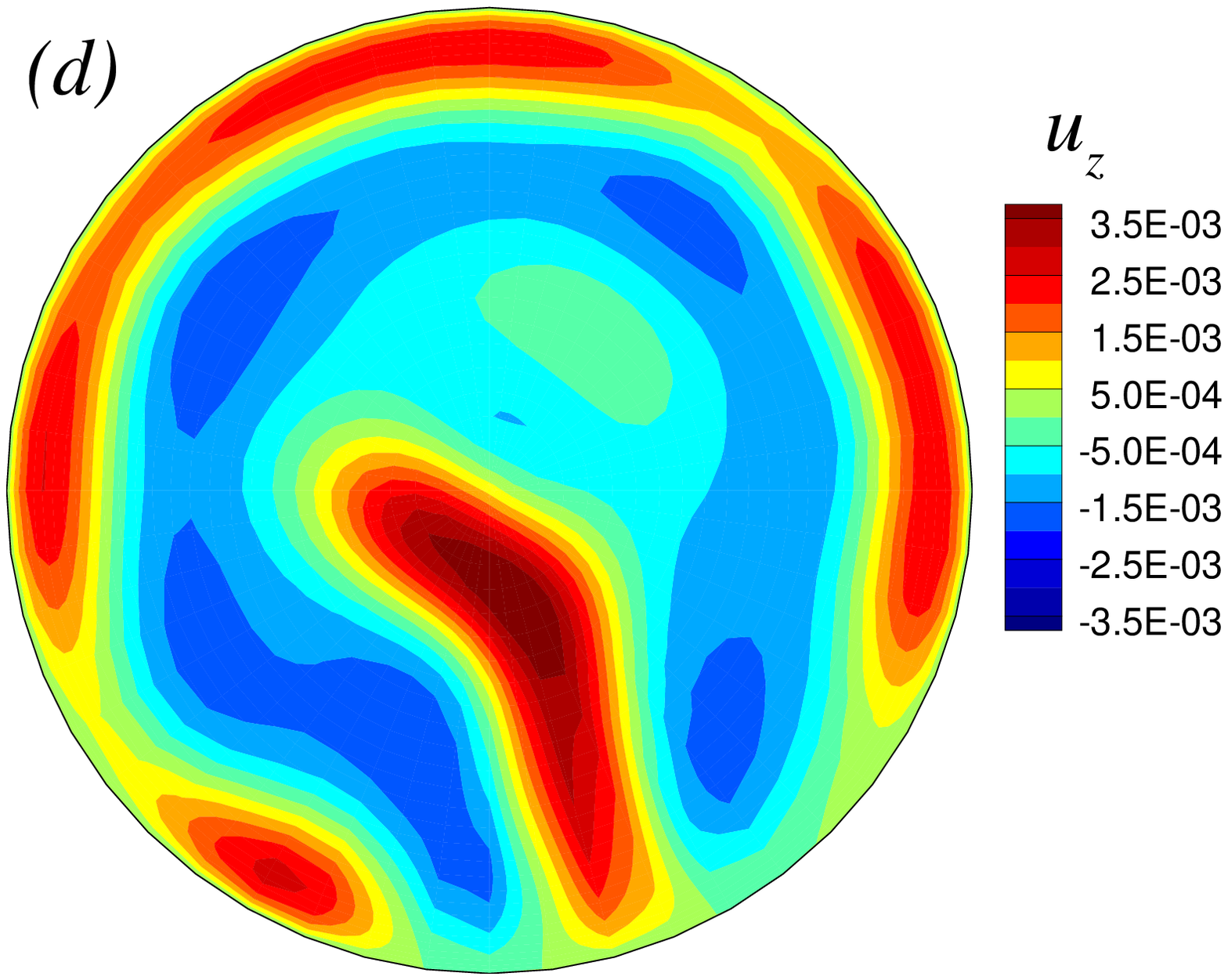}
\caption{Spatial structure of developed finite-amplitude convection flow obtained in the simulation with $H_E=0.58$ and $\Gras=5.0\times 10^5$ at $t=600$. \emph{(a)}, Full temperature  $T$ and velocity vectors (drawn at every second point in each direction) in the vertical cross-section $\theta=0$, $\pi$. \emph{(b)}--\emph{(d)}, vertical velocity in the horizontal cross-sections $z=0.25$  \emph{(b)}, $z=0$ \emph{(c)}, and $z=0.75$ \emph{(d)}.
}
\label{fig:sp3}
\end{center}
\end{figure}

\begin{figure}
\begin{center}
\includegraphics[width=0.5\textwidth]{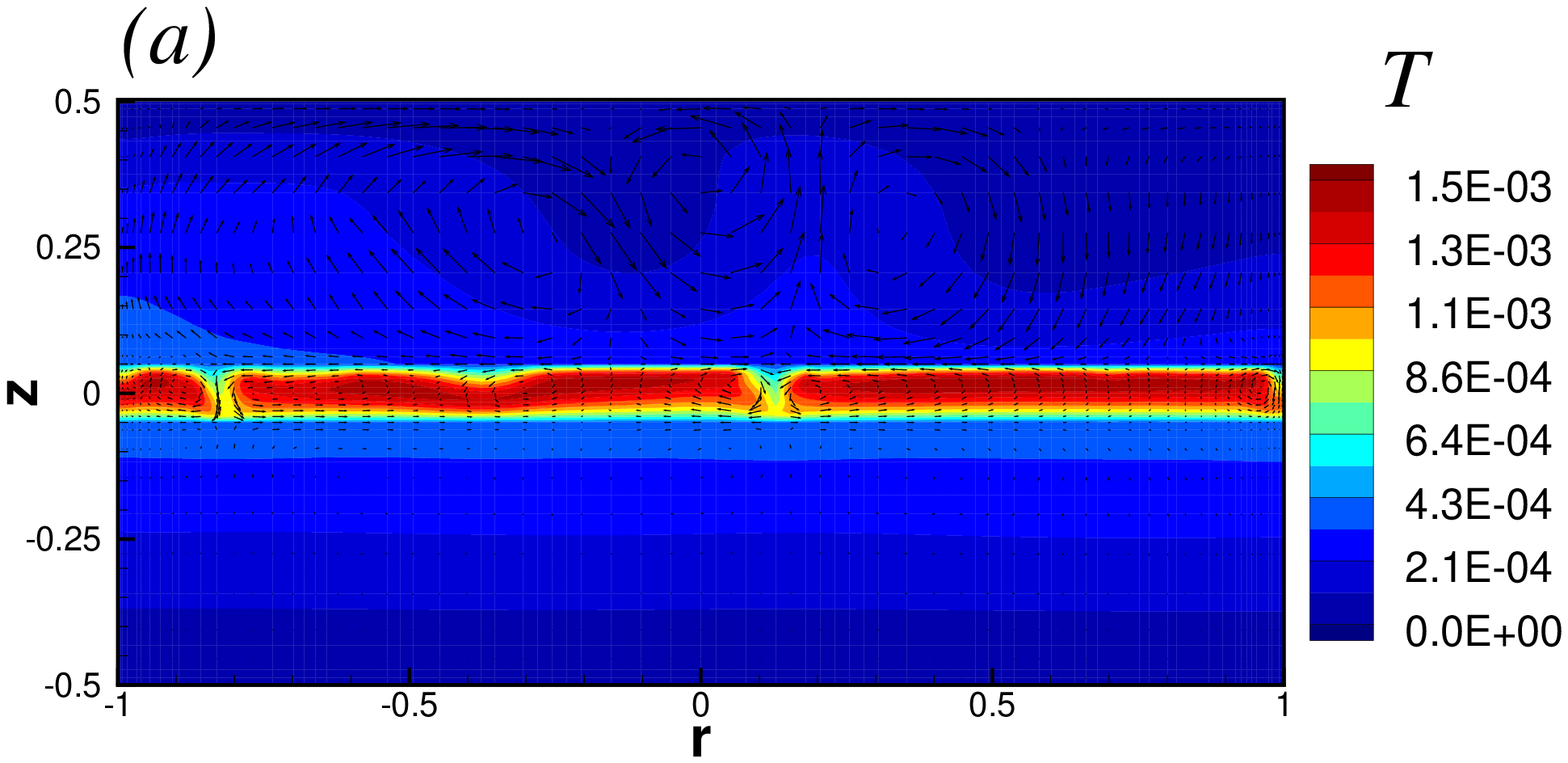}\includegraphics[width=0.5\textwidth]{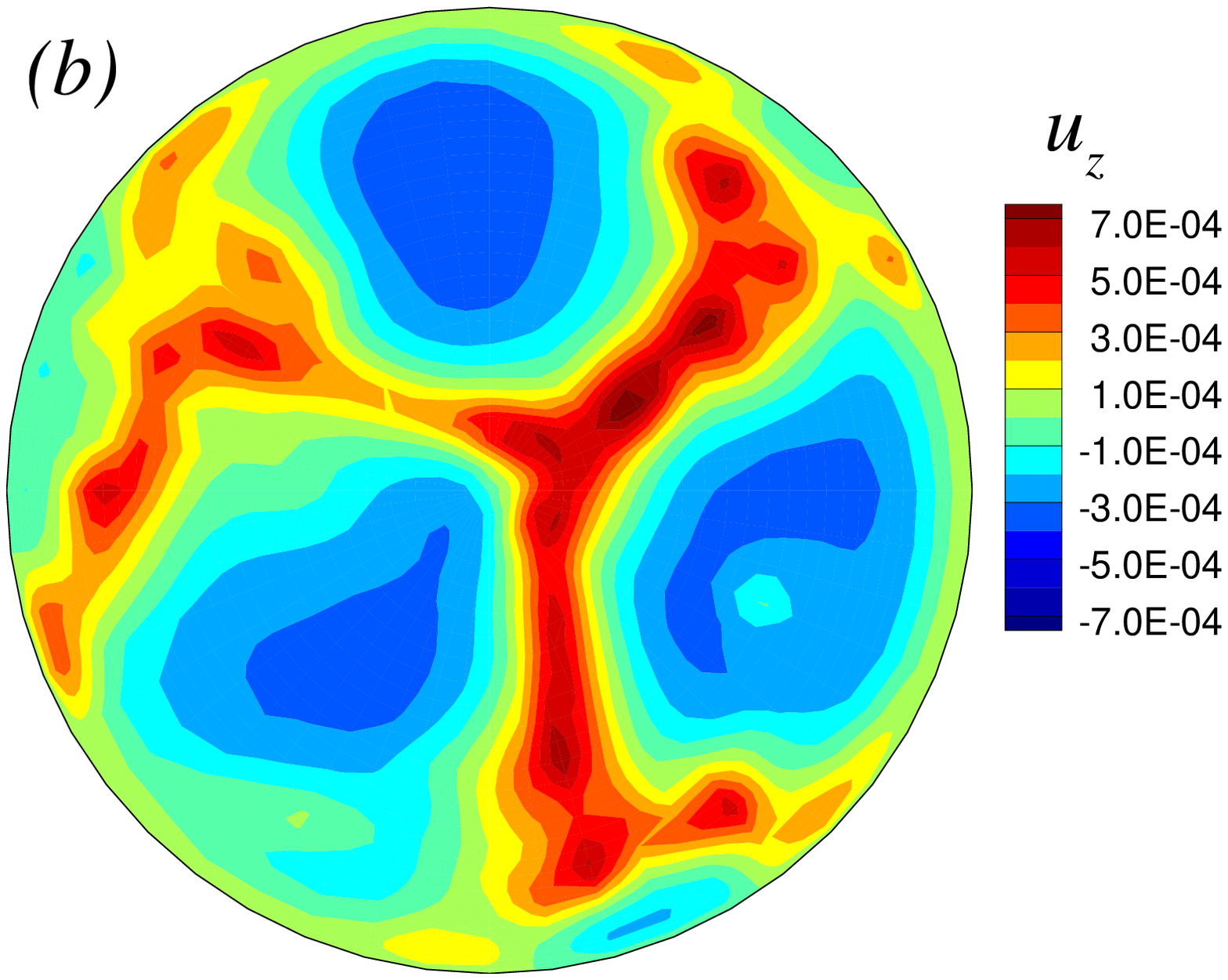}\\
\includegraphics[width=0.5\textwidth]{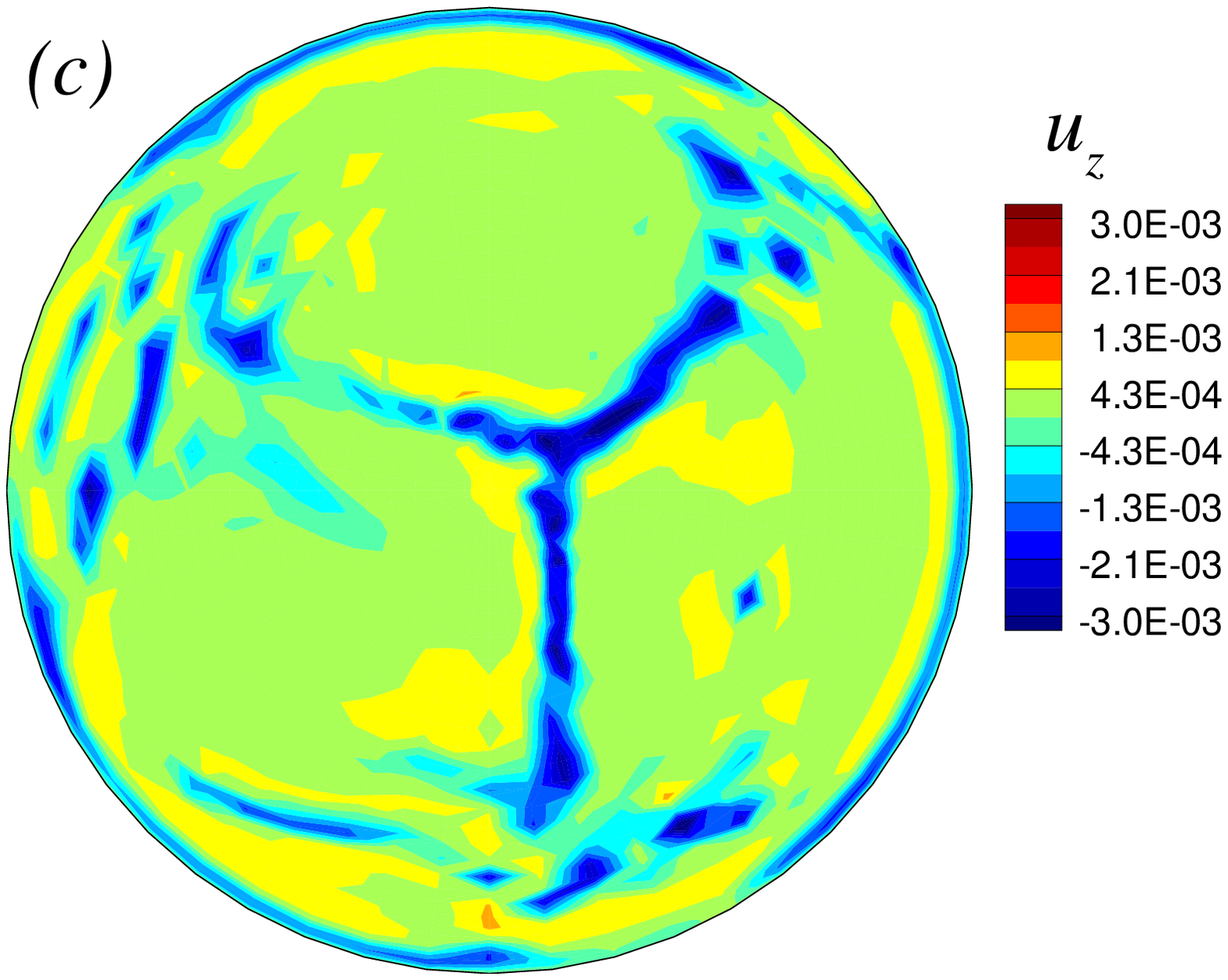}\includegraphics[width=0.5\textwidth]{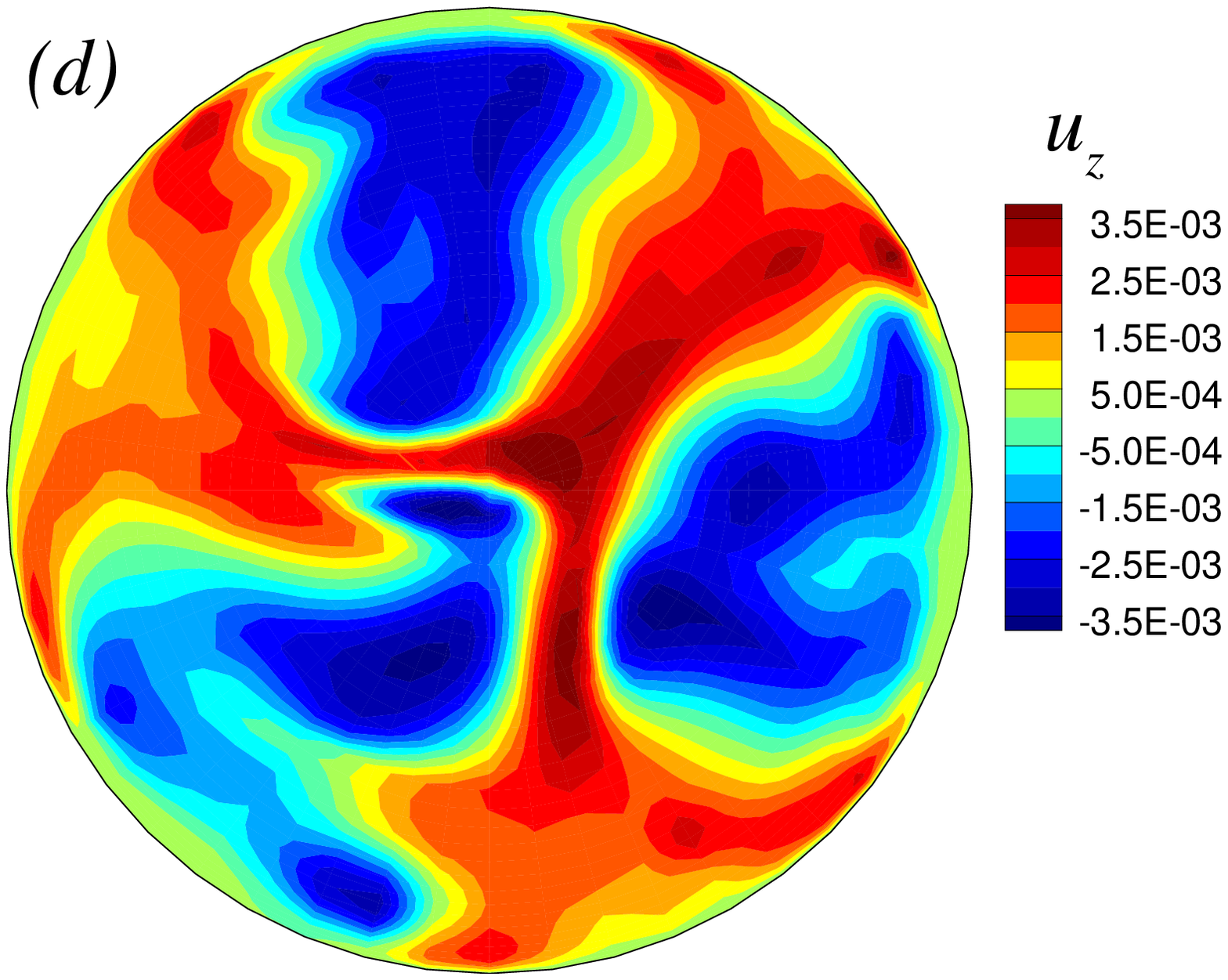}
\caption{Spatial structure of developed finite-amplitude convection flow obtained in the simulation with $H_E=0.1$ and $\Gras=3.0\times 10^9$ at $t=4488$. \emph{(a)}, Full temperature  $T$ and velocity vectors (drawn at every second point in each direction) in the vertical cross-section $\theta=0$, $\pi$. \emph{(b)}--\emph{(d)}, vertical velocity in the horizontal cross-sections $z=0.25$  \emph{(b)}, $z=0$ \emph{(c)}, and $z=0.75$ \emph{(d)}.
}
\label{fig:sp4}
\end{center}
\end{figure}

The typical structures of the fully developed convection flows in batteries with $H_E=0.58$ and $H_E=0.1$ are shown, respectively, in figures \ref{fig:sp3} and \ref{fig:sp4}. We see that the principal flow structure remains the same as in the case of $H_E=0.304$. It can be described as a pattern of irregular three-dimensional comvection cells. Some variations are observed, though. In particular, the flow at  $H_E=0.58$ is characterized by a smaller number of larger convection cells. The flow at $H_E=0.1$ demonstrates a flow pattern dominated by sheets of strong upward or downward velocity similar to the patterns often found in convection in shallow layers (see, e.g. \cite{Zikanov:2002} for an example of such a pattern in the case of penetrative convection).

Figures \ref{fig:sp2}--\ref{fig:sp4} and table \ref{table1} show that the partition of the kinetic energy of the flow among the three layers changes with $H_E$. At $H_E=0.58$, the strongest flow is still generated in the electrolyte layer, the flows in the layer \textsf{A} and, especially, layer \textsf{B} being much weaker. The situation changes at $H_E=0.1$, when the flow in the layer \textsf{A} has the average kinetic energy about an order of magnitude higher than the flow in the layer \textsf{E}. The flow induced in the bottom layer \textsf{B} remains weak.

\section{Discussion}
\label{sec:disc}
The only plausible explanation of the convection instability reported in section \ref{sec:res} is the unstable stratification of the base temperature in the upper part of the electrolyte layer (see fig.~\ref{fig:geom}b). 
This is indicated by the fact that the unstable temperature gradient is much higher in the electrolyte layer than in the layer \textsf{A} and by the structure of the instability mode illustrated in figure \ref{fig:sp1}. 

It is interesting to explore the analogy between our case and the convection instability in a uniformly internally heated layer between two horizontal walls (see, e.g. \cite{Kulacki:1972}) or the classical Rayleigh-B\'{e}nard instability in a layer between stress-free boundaries \cite{Chandrasekhar:1961}. For that purpose, we 
use the dimensional height of the unstably stratified sub-layer $H_E^*/2$ as the typical length scale and
redefine the  temperature scale as the typical difference between the maximum and minimum base-state temperature in the electrolyte layer $\Delta T^*=Q_0\left(H^*_E\right)^2/8\kappa_E$.  This results in the effective Rayleigh number defined in terms of our non-dimensional parameters as
\begin{equation}
Ra^{eff}\equiv \frac{g\beta_E\rho_EC_E}{\nu_E\kappa_E}\Delta T^*\left(\frac{H^*_E}{2}\right)^3=\frac{g\beta_E \rho_E C_E Q_0 (H^*_E)^5}{64\nu_E\kappa_E^2}=\frac{1}{64}\Pran \Gras H_E^5.
\label{Raeff}
\end{equation}
Recalculating from table \ref{table2}, we find the instability thresholds $Ra^{eff}_{cr}=241$ at $H_E=0.1$, $Ra^{eff}_{cr}=305$ at $H_E=0.304$, and $Ra^{eff}_{cr}=785$ at $H_E=0.58$. The values  are of the same order of magnitude as the analogously defined $Ra^{eff}_{cr}=560$ for the internally heated layer \cite{Kulacki:1972} and $Ra^{eff}_{cr}=27\pi^4/4\approx 657$ for the Rayleigh-B\'{e}nard convection with stress-free boundaries \cite{Chandrasekhar:1961}. 

Quantitatively, however, our values are different. Of particular concern is the fact that the threshold found for the shallow layer $H_E=0.1$ is about two times smaller than in \cite{Kulacki:1972} and \cite{Chandrasekhar:1961}. This can be explained bt 
 the differences in the system's geometry. The zero shear stress conditions imposed on both boundaries in the Rayleigh-B\'{e}nard case or the no-slip condition on the upper boundary in the internal heating case restrict development of convection rolls in the unstably stratified layer and, thus, delay the instability in comparison to our case, in which freely movable liquid above and below the unstably stratified layer create `soft' boundary conditions. In order to validate this explanation, we have conducted additional simulations with $H=0.1$ and $H_E=0.09$, i.e. for a cell, in which the layers \textsf{A} and \textsf{B} are very thin and the system approaches the single internally heated layer analyzed in \cite{Kulacki:1972}. The instability threshold $Ra^{eff}_{cr}\approx 548$, much closer  to the results of  \cite{Kulacki:1972}, was found.

\begin{figure}
\begin{center}
\includegraphics[width=0.7\textwidth]{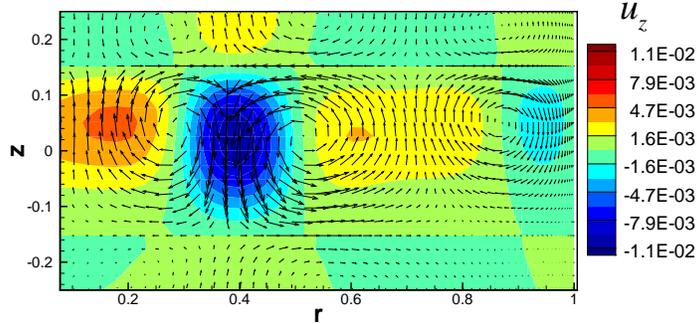}
\caption{Spatial structure of developed finite-amplitude convection flow obtained in the simulation with $H_E=0.304$ and $\Gras=4.5\times 10^6$ at $t=1500$.  Vertical velocity  $u_z$ and velocity vectors (drawn at every grid point) are shown in a part of the vertical cross-section $\theta=0$, $\pi$. }.
\label{fig:sp5}
\end{center}
\end{figure}

Flows in the liquid metal layers \textsf{A} and \textsf{B} are induced by the flow developing in the electrolyte via interfacial coupling. The two major coupling mechanisms allowed by our model are the viscous shear stress and  conduction heat transfer across the interface\footnote{There is also electromagnetic coupling, but it is, as we  discuss later in this section, is weak.}. The shear stress is active at both the interfaces. It leads to formation of  circulation cells in the layers \textsf{A} and \textsf{B} that have the same direction of horizontal velocity and, thus, the opposite circulation sign as the adjacent convection cell in the electrolyte. In such a flow, the zones of upward (downward) velocity in the layer \textsf{E} should approximately correspond to zones of downward (upward) velocity in layers \textsf{A} and \textsf{B}. 

The coupling by heat conduction is active only at the \textsf{A}-\textsf{E} interface, where upward flows of electrolyte create hot spots, which enhance the unstable stratification in the layer \textsf{A} and may cause convection there. Considered in isolation from other coupling mechanisms, this should cause an upward flow in the layer \textsf{A}   above the  zones of upward flow in the layer \textsf{E}.

Visual inspection of the flow structures in figures \ref{fig:sp2}-\ref{fig:sp4} as well as of the additional illustration in figure \ref{fig:sp5} suggests that the shear stress is the predominant mechanism of interfacial coupling in our system. The heat conduction  across the interface plays, at best, a secondary role. We note that this conclusion as well as the observed general structure of the flow can change when deformation of interface, surface tension, and other so far neglected factors of the system's dynamics are included  into the model.

In accordance with the mechanisms of flow generation, the amplitude of convection flow varies among the three layers. It is expected that the primary buoyancy-driven flow in the electrolyte layer is stronger than the secondary flows in the layers \textsf{A} and \textsf{B}. Table \ref{table1} confirms this, but only for the cases of thick ($H_E=0.58$) and intermediate ($H_E=0.304$) electrolyte layers. In the case of a thin layer ($H_E=0.1$), the flow in the layer \textsf{A} has, on average, stronger velocity than the flow in the layer \textsf{E}. This can be explained by the restrictive action of close top and bottom boundaries of the layer \textsf{E}, which prevents development of large convection cells in this layer.

The convection introduces mixing that affects the heat transfer through the battery. In the framework of our model with insulating sidewalls and top and bottom walls kept at the same temperature, the effect is such that the heat transfer through the top wall increases, while the heat transfer through the bottom wall decreases by approximately the same amount in the time-averaged sense (see the values of $\Nu_A$ and $\Nu_B$ in table \ref{table1}). The  temperature peak within the electrolyte layer is suppressed, but not dramatically so (see figure \ref{fig:nusselt}b). We conclude that the convection has some, but not very strong effect on the temperature regime of the battery. The conclusion may change when significantly higher $\Gras$ are considered.

Our approach in this study is to take into account the magnetohydrodynamic (MHD) effects in the form of the Lorentz force $\Ha^2\Rey^{-1}\bm{j}\times \bm{B}_0$ arising from the interaction between the electric currents induced by the flow and the base magnetic field $\bm{B}_0$. This is done to account for possible effects of the force on the flow, in particular, for suppression of the unstable convection modes, delay of the instability, and formation of anisotropic flow structures elongated in the direction of $\bm{B}_0$ (see, e.g., \cite{Davidson:2001} for an introductory discussion and \cite{Krasnov:2012,Zhao:2012,ZikanovJFM:2013,Zhang:2014,Zikanov:1998} for examples of these effects). The results presented so far in this paper, however, do not demonstrate clearly visible MHD effects. In particular, the anisotropy, which would, in our case, take the form of tendency to axially symmetric flow structures, is not seen. 

An explanation to the absence of strong MHD effects is found when we calculate the Stuart number 
\begin{equation}
\label{stuart}
\N=\frac{\Ha^2}{\Rey}=\frac{B_0^2R\sigma_E}{\rho_E U},
\end{equation}
which is an estimate of the typical ratio between the Lorentz and inertial forces acting in the flow and is typically taken as a measure of the anticipated degree of the MHD transformation of the flow. Two corrections are needed to find the relevant value of $\N$. First, since the MHD effects are much stronger in highly conducting metals \textsf{A} and \textsf{B}, conductivity $\sigma_A=\sigma_B=10^4\sigma_E$ must be used. Second, we see in figures \ref{fig:sp2}--\ref{fig:sp4} and table \ref{table1} that the non-dimensional velocity of the developed convection flow has small (about $10^{-2}$ or smaller) amplitude of velocity. This is related to the known fact that the 
free-fall velocity $U$ (see (\ref{scales})) overestimates the actual velocity scale in a convection flow. Using $U$ in (\ref{stuart}) leads to underestimating $\N$ and, thus, the strength of the MHD effects. 

The effective Stuart number can, therefore, be computed as (we use the expressions (\ref{reynolds}), (\ref{hagra}), (\ref{values}) to compute $\Ha$ and $\Rey$)
\begin{equation}
\label{stuarteff}
\N_{eff}\approx10^2 \frac{\sigma_A}{\sigma_E} \N = 10^6 \beta^2 \Gras^{3/10} = 1.1025\times 10^{-6} \Gras^{3/10}.
\end{equation}
Finally, using the instability threshold  values $\Gras_{cr}$ in table \ref{table2} we find $\N_{eff}\approx 4.4\times 10^{-4}$ at $H_E=0.1$, $ 8.9\times 10^{-5}$ at $H_E=0.304$, and $4.48\times 10^{-5}$ at $H_E=0.58$. We see that $\N_{eff}\ll 1$ and, so, the effect of the magnetic field on the velocity field of the convection flow is negligible (see, e.g., \cite{Zikanov:1998}).

As an approximation, we can take the value $N_{eff}=0.1$ as the one, at which the MHD effects become significant. According to (\ref{stuarteff}), this corresponds, in the framework of our model, to $\Gras=3.35\times 10^{16}$.

\section{Implications for battery design}\label{sec:impl}
We start with the estimates of the critical radius $R_{cr}$ such that the batteries of larger radii experience the convection instability. Using the values of $\Gras_{cr}$ in table \ref{table2} and the formula (\ref{estimateR}) we find $R_{cr}=3.6$ cm at $H_E=0.1$, $R_{cr}=1.2$ cm at $H_E=0.304$, and $R_{cr}=0.78$ cm at $H_E=0.58$. 

We can also estimate the expected intensity of the convection flow. Particularly interesting is the flow in the bottom layer. As discussed in \cite{Kim:2013,Kelley:2014}, this flow plays a potentially beneficial role, as it enhances the mixing of metal \textsf{B} and compound \textsf{A}-\textsf{B}, thus reducing a limiting factor of the reaction rate. The typical value of the non-dimensional velocity can be evaluated from table \ref{table1} as $U_B\sim E_B^{1/2}$. Multiplying by the velocity scale $U=\nu_E\Gras^{1/2}R^{-1}$ see (\ref{scales}) and using the physical properties from section \ref{sec:parameters} and current $J_0=1$ A/cm$^2$, we find the typical dimensional amplitude of velocity in the bottom layer as
\begin{equation}
\label{ub-est}
U_B^{dim}\sim 10^{-3}E_B^{1/2}\Gras^{3/10} \textrm{ m/s}.
\end{equation}
This gives rather low values of velocity. For example, in the particularly interesting case of thin electrolyte layer $H_E=0.1$ at $\Gras =3\times 10^9$, we find $U_B^{dim}\sim $ 0.3 mm/s. Even lower values are considered in the other cases considered in our study. We must note, however, that the example above corresponds to a small cell (radius about 5.2 cm with the current and physical parameters used in our work). Much stronger circulation and turbulent flow are expected in larger batteries. For small batteries, desired mixing can be produced by other mechanisms, such as bottom heating, electrovortex instability, etc.

We can make preliminary estimate of the interface deformation cause by the convection flow. The question is interesting for two reasons. First, the possibility of significant deformation would invalidate our model, based on the assumption of non-deformable interfaces. Second, a deformation so string as to cause rupture of the electrolyte layer at some point is highly undesirable, since this would lead to short circuit between the metal layers and, thus, a disruption of the battery's operation. The effect cannot be analyzed directly within the framework of our model, but can be approached via a qualitative energy arguments similar to those used for the Tayler instability in \cite{Herreman:2015}. We recall that the densities of the melts in a real battery are strongly different and consider the practically most interesting case of thin electrolyte layer.
The estimate is based on comparison between the kinetic energy of the flow with the gravitational potential of the interface deformation. Since the density $\rho_B$ of the melt in the bottom layer \textsf{B} is much higher than $\rho_E$ and $\rho_A$ and since the flow is the strongest in the layer \textsf{A}, we only consider the deformation of the upper interface. A displacement of a fluid particle at the interface by the dimensional distance $h^*$ increases the specific gravitational potential by 
\begin{equation}
\label{gravpot}
E_{pot}=\left(\rho_E-\rho_A\right)gh^*,
\end{equation}
which should be of the same order of magnitude as the specific kinetic energy of the flow
\begin{equation}
\label{kinen}
E_{kin}=\frac{1}{2}\rho_AU^{*2},
\end{equation}
where $U^*$ is the dimensional mean velocity in the layer \textsf{A}.  This leads to the estimate
\begin{equation}
\label{hstar}
h^*\sim \frac{1}{2}\frac{\rho_A}{\rho_E-\rho_A}\frac{U^{*2}}{g}
\end{equation}
or, in the non-dimensional form
\begin{equation}
\label{hnon}
h\equiv \frac{h^*}{H^*_E}\sim \frac{1}{2}\frac{\rho_A}{\rho_E-\rho_A}Fr^2,
\end{equation}
where $H^*_E$ is the dimensional thickness of the undisturbed electrolyte layer and
\begin{equation}
\label{froude}
\Fr=\frac{U^*}{\left(gH^*_E\right)^{1/2}}
\end{equation}
is the Froude number. 

The event of short circuit between the metal electrodes corresponds to $h^*\ge H^*_E$, i.e. the Froude number exceeding the critical value 
\begin{equation}
\label{frcr}
\Fr_{cr}=\left(2\frac{\rho_E-\rho_A}{\rho_A}\right)^{1/2}.
\end{equation}

The typical values of the coefficient $\left(\rho_E-\rho_A\right)/\rho_A$ is about 2. We consider the case of $H_E=0.1$ and $\Gras=3\times 10^9$, where the strongest flow is found above the thinnest electrolyte layer, and estimate the square of the mean velocity as $U^{*2}\sim U^2 E_A$, where $U$ is the free-fall velocity (\ref{scales}) and $E_A=2.76\times 10^{-5}$ is the non-dimensional kinetic energy density from  table \ref{table1}. This gives
\begin{equation}
\label{frest}
\Fr^2\sim E_A\frac{U^2}{gH^*_E}=\frac{E_A}{H_E}\frac{\alpha_EJ_0^2R^2}{\sigma_E\kappa_E}.
\end{equation}
With the physical parameters used in our study, $\Gras=3\times 10^{9}$ corresponds to the battery radius $R=5.2$ cm, which gives $\Fr^2\sim 3\times 10^{-4}$. We see that the amplitude of the deformation of the interface is about four orders of magnitude smaller than the thickness of the electrolyte layer. The validity of our modeling assumption is confirmed, and the danger of the rupture of interface by the convection flow should be seen as immaterial. These conclusions are, of course, only valid for the range of parameters considered in our paper. The situation can be completely different in larger battery cells.

Finally, we estimate the size of the battery, at which the convection flow is significantly affected by the magnetic field $\bm{B}_0$. Following the calculations at the end of the previous section we find, for $\Gras=3.35\times 10^{16}$, the radius $R=1.33$ m. At smaller sizes, the effect can be ignored in the analysis. At the same time, attention may be needed to the effects of the magnetic fields generated externally, for example, by the current supply lines.

\section{Concluding remarks}\label{sec:concl}
The main conclusion of the analysis presented in this paper is that thermal convection is virtually inevitable in liquid metal batteries. It is present in even small-scale prototypes. In the practically most interesting case of thin electrolyte layer, the flow is the strongest in the top layer containing the liquid metal and the weakest in the bottom layer. 

The effect of the flow on the battery's operation is found to be weak. The velocity in the bottom metal layer is too small to generate strong mixing of reactants. No significant deformation of the electrolyte-metal interfaces is expected.

These conclusions are, however, only valid for small (a few cm in radius) batteries, for which computations have been conducted in this study. The effect can and should be anticipated to be mush stronger in larger batteries, for which the Grashof number $\Gras\sim R^5$ is many orders of magnitude larger. 

It appears that the most interesting direction of future work is the analysis of operation of large ($R\sim$ 0.1-1 m) batteries. In addition to being computationally challenging, this will require more complex physical models, which will include deformation of interfaces, dynamics of the base current distribution $J_0$, etc. Ultimately, it will be necessary to develop a comprehensive model that describes the multiple mechanisms of instability and flow generation as coupled physical phenomena.

\paragraph{Acknowledgements}
Financial support is provided by the U.S. National Science Foundation (Grant CBET 1435269) and by the University of Michigan - Dearborn.

\bibliographystyle{elsarticle-num} 
\bibliography{../../battery}





\end{document}